\documentclass[a4paper,11pt]{article}
\PassOptionsToPackage{breaklinks}{hyperref}
\usepackage{jheppub}

\usepackage{graphicx}
\usepackage{dcolumn}
\usepackage{bm}
\usepackage{hyperref}
\usepackage[dvipsnames]{xcolor}
\usepackage{amsmath}
\usepackage{textpos}
\usepackage{physics}
\usepackage{siunitx}
\usepackage{float}
\usepackage[T1]{fontenc}
\usepackage{xspace}
\newcommand{\figwid}{0.65}
\newcommand{\twofigwid}{0.49}
\newcommand{\pythia}{\textsc{pythia}\xspace}
\newcommand{\nuwro}{\textsc{NuWro}\xspace}
\newcommand{\pn}{p_\textrm{N}}
\newcommand{\wth}{W_\text{th}}
\newcommand{\wmin}{W_\text{min}}
\newcommand{\wmax}{W_\text{max}}
\newcommand{\diff}{\text{d}}
\newcommand{\sigmaSPP}{\sigma^\text{SPP}}
\newcommand{\sigmaRES}{\sigma^{\Delta}}
\newcommand{\sigmaPySPP}{\sigma^\text{\pythia:SPP}}

\usepackage{booktabs}
\usepackage{gensymb}
\usepackage{multirow}
\usepackage{lipsum}
\usepackage[normalem]{ulem}

\usepackage[caption=false]{subfig}
\newcommand{\figref}[2][{}]{\hyperref[#2]{\figurename~\ref{#2}#1}}

\newcommand{\warwick}{University of Warwick, Department of Physics,\\ Coventry, CV4 7AL United Kingdom}
\newcommand{\ucas}{School of Physical Sciences, University of Chinese Academy of Sciences,\\ Beijing 100049, China}

\begin{document}

\title{The Ghent Hybrid Model in \nuwro: a new neutrino single-pion production model in the GeV regime}

\author[a,b]{Qiyu Yan,}
\author[c]{Kajetan Niewczas,}
\author[d,c]{Alexis Nikolakopoulos,}
\author[e,f]{Raúl González-Jiménez,}
\author[c]{Natalie Jachowicz,}
\author[b]{Xianguo Lu,}
\author[g]{Jan Sobczyk,}
\author[a]{and Yangheng Zheng}
\affiliation[a]{\ucas}
\affiliation[b]{\warwick}
\affiliation[c]{Department of Physics and Astronomy, Ghent University,\\ Proeftuinstraat 86, B-9000 Gent, Belgium}
\affiliation[d]{Theoretical Physics Department, Fermilab,\\ Batavia IL, USA}
\affiliation[e]{Grupo de Física Nuclear, Departamento de Estructura de la Materia, Física Térmica y Electrónica, Facultad de Ciencias Físicas, Universidad Complutense de Madrid and IPARCOS}
\affiliation[f]{Departamento de Física Atómica, Molecular y Nuclear, Universidad de Sevilla,\\ 41080 Sevilla, Spain}
\affiliation[g]{Institute of Theoretical Physics, University of Wroc\l aw,\\ pl. M. Borna 9,
    50-204, Wroc\l aw, Poland}

\date{\today}

\abstract{Neutrino-induced single-pion production constitutes an essential interaction channel in modern neutrino oscillation experiments, with its products building up a significant fraction of the observable hadronic final states. Frameworks of oscillation analyses strongly rely on Monte Carlo neutrino event generators, which provide theoretical predictions of neutrino interactions on nuclear targets. Thus, it is crucial to integrate state-of-the-art single-pion production models with Monte Carlo simulations to prepare for the upcoming systematics-dominated landscape of neutrino measurements. In this work, we present the implementation of the Ghent Hybrid model for neutrino-induced single-pion production in the \nuwro Monte Carlo event generator. The interaction dynamics includes coherently-added contributions from nucleon resonances and a non-resonant background, merged into the \pythia branching predictions in the deep-inelastic regime, as instrumented by \nuwro. This neutrino-nucleon interaction model is fully incorporated into the nuclear framework of the generator, allowing it to account for the influence of both initial- and final-state nuclear medium effects. We compare the predictions of this integrated implementation with recent pion production data from accelerator-based neutrino experiments. The results of the novel model show improved agreement of the generator predictions with the data and point to the significance of the refined treatment of the description of pion-production processes beyond the $\Delta$ region.}

\maketitle

\section{Introduction}\label{intro}

In the energy range of a few GeV, as explored by accelerator-based neutrino experiments such as DUNE~\cite{DUNE:whitepaper}, Hyper-Kamiokande~\cite{HK}, NOvA~\cite{nova}, and T2K~\cite{ABE2011106}, the understanding of the neutrino-nucleus scattering cross section is limited to a precision of, at best, 10\%~\cite{DUNE:whitepaper}, corresponding to a 3-7\% contribution to the overall experimental systematic error~\cite{T2K-naturepaper}. Alongside the normalization of the neutrino flux, this limitation represents a significant source of uncertainty in oscillation analyses. Achieving the measurement of the CP violation phase, which can shed light on mechanisms explaining the prevalence of matter over antimatter in the current universe, requires reducing systematic errors to as low as 1\%~\cite{DUNE:whitepaper}.
Neutrino-induced single-pion production (SPP) is an important reaction channel in these experiments.
Its significance extends to atmospheric neutrino programs in Super-Kamiokande~\cite{Super-Kamiokande:1998kpq}, Hyper-Kamiokande, and JUNO~\cite{juno}, where GeV neutrinos play an important role in determining the neutrino mass ordering.

We will focus on discussing SPP in the \nuwro Monte Carlo (MC) generator. \nuwro is one of the major MC generators extensively used in studies by experimental groups, and the accuracy of its predictions is vital for the precision of neutrino oscillation measurements.
Pion production in neutrino interactions occurs via two primary mechanisms: resonance production and non-resonance production.
The current \nuwro implementation utilizes a dedicated $\Delta$ resonance production and decay model for the resonant contribution. The final state angular distribution is taken from the experimental results from ANL or BNL~\cite{Graczyk}. This model accurately describes the inclusive cross section for pion production through $\Delta$ resonance formation~\cite{Graczyk}. However, the reliability of the \nuwro approach decreases in the higher invariant mass ($W$) region due to the absence of explicit contributions from other resonances.

To address this limitation, \nuwro incorporates \pythia~\cite{pythia6} to model the final state originating from other resonances within the framework of non-resonant production, using inclusive cross sections from the Bodek-Yang approach of quark parton distribution functions~\cite{Bodek:2003wd}. Additionally, \pythia is employed in the $\Delta$ region to account for pion production through non-resonant processes. However, using Bodek-Yang approach and \pythia hadronization models at low $W$ and low squared-four momentum transfer ($Q^2$) raises concern. In this kinematic regime, interactions do not primarily occur on quarks, and the underlying assumptions of the Lund model, the theoretical foundation of \pythia, are barely met~\cite{collins2004factorizationhardprocessesqcd}.

\begin{figure}[!tb]
    \centering
    \includegraphics[width=\figwid\textwidth]{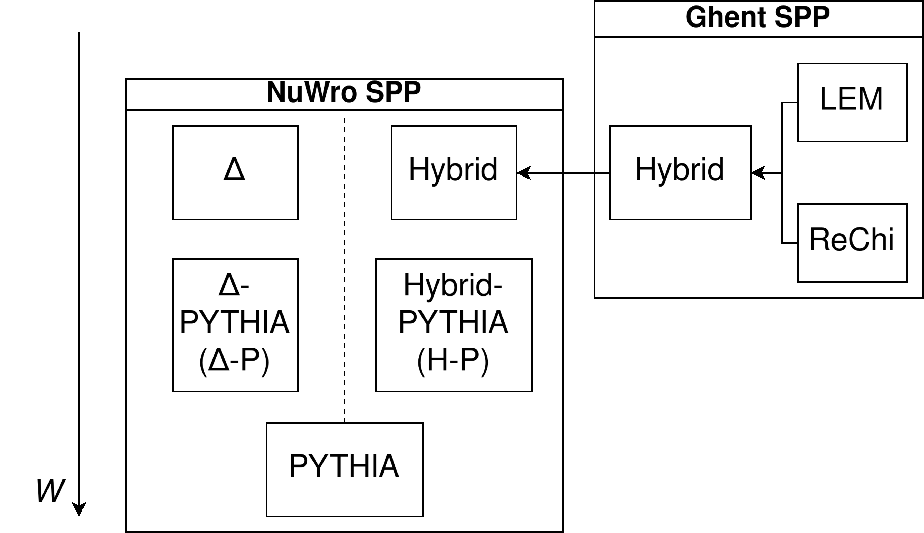}
    \caption{Diagrammatic definitions of the $\Delta$-\pythia ($\Delta$-P) and Hybrid-\pythia (H-P) models, with the former combining the $\Delta$ model and \pythia in \nuwro, while the latter replaces the $\Delta$ model with the Ghent SPP Hybrid model. All models are arranged by increasing $W$, as indicated by the long arrow.
    }
    \label{fig:blueprint}
\end{figure}

To address these drawbacks, in Ref.~\cite{Niewczas21}, different approaches to implement in MC generators the model of Refs.~\cite{Gonzalez-Jimenez17, Nikolakopoulos23} for SPP off the nucleon were discussed. The idea is to model neutrino-induced SPP across the entire kinematic range, extending up to potentially unlimited $W$ values. Consequently, there is no reliance on the hadronization model for the SPP channel, enabling modeling of the differential cross section across the complete kinematic spectrum.
In this study, we incorporate this methodology into the
\nuwro event generator framework, ensuring its seamless integration with the nuclear model and the intranuclear cascade.
To evaluate its efficacy, this new model implementation is compared against existing datasets. Specifically, we will utilize the MINERvA $\pi^+$ sample~\cite{MINERvA:2014ogb} and MicroBooNE $\pi^0$ data~\cite{microboone:pi0}, which are mostly sensitive to the $\Delta$ resonance region, and the MINERvA $\pi^0$ transverse kinematic imbalance (TKI)~\cite{MINERvA:2020anu} data, which has significant contributions from heavier resonances.

The paper is structured as follows. In Section~\ref{Hybrid-model} we summarize the main features of the Ghent Hybrid model. Section~\ref{nuwro} details the default description of SPP processes in \nuwro and the implementation of the transition to the regime where multi-pion and deep inelastic scattering (DIS) events become important, both in standard \nuwro and in the new implementation incorporating the Hybrid model (Fig.~\ref{fig:blueprint}). The resulting theory predictions are compared to the experimental measurements in Sec.~\ref{datacomparison}, followed by the conclusions in Sec.~\ref{conclusion}.

\section{The Ghent Hybrid model for single-pion production}\label{Hybrid-model}

The Hybrid model~\cite{Gonzalez-Jimenez17} for electroweak SPP developed by the Ghent group consists of two main components, each targeting distinct kinematic regions.  The low-energy part of the model (LEM) includes contributions from resonances and background based on tree-level diagrams described in Ref.~\cite{Hernandez07}. It has been systematically benchmarked against data and the MAID07~\cite{MAID07} and dynamical coupled channels models~\cite{Kamano16, Nakamura15} with satisfactory agreement~\cite{Gonzalez-Jimenez17,PhysRevD.98.073001,Nikolakopoulos23}. For the $\Delta$ resonance, the vector form factors of Ref.~\cite{Lalakulich06} are used. The axial form factors were determined through analysis of bubble chamber data detailed in Ref.~\cite{Alvarez-Ruso16}. The vector form factors for the higher mass resonances ($P_{11}(1440)$, $D_{13}(1520)$, $S_{11}(1535)$) are described in Ref.~\cite{Nikolakopoulos23}, drawing upon findings from Refs.~\cite{MAID07, Hernandez08, Lalakulich06}. The axial form factors for higher-mass resonances are reported in Ref.~\cite{Gonzalez-Jimenez17}. At $Q^2 = 0$, they are determined based on  partially conserved axial current (PCAC) considerations, disregarding any undetermined axial couplings. The $Q^2$-dependence of the axial form factors is taken from Ref.~\cite{Lalakulich06}.

At high energies ($W$ above $1.5$-$2$ GeV), the LEM model starts exhibiting anomalous behavior because it relies solely on tree-level diagrams. To overcome this limitation, the Hybrid model employs a description of the non-resonant background based on Regge phenomenology (ReChi). In this approach, the tree-level propagator of $t$-channel meson exchanges in the low-energy background is replaced by a Regge propagator~\cite{Guidal97}. This approach offers an efficient description of the forward-scattering process, which largely dominates in the high-energy regime~\cite{Guidal97, Kaskulov:Mosel, Vrancx14a}. Additionally, the high-energy behavior of the resonances is regulated by the inclusion of cut-off hadronic form factors~\cite{Vrancx11}.

The transition between the low- and high-energy background is implemented, at the amplitude level, in a phenomenological way as a function of $W$:
\begin{equation}
    J^\mu_\text{Hybrid} = J^\mu_{\text{RES}} + \cos^2\phi(W)J^\mu_\text{LEM} + \sin^2\phi(W)J^\mu_\text{ReChi},
\end{equation}
where $J^\mu_\text{Hybrid}$ is the full hadronic current, while $J^\mu_\text{RES}$, $J^\mu_\text{LEM}$, and $J^\mu_\text{ReChi}$ denote the contributions from resonances, low-energy background, and high-energy background, respectively.
The transition function $\phi(W)$ depends on $W$ and is defined as follows:
\begin{equation}
    \phi(W ) = \frac{\pi}{2}\left[ 1-\frac{1}{1+\exp\left(\frac{W-W_0}{L}\right)}\right].
\end{equation}
Here, $W_0$ and $L$ represent the center and width of the transition, respectively, fixed at $\SI{1.5}{~GeV}$ and $\SI{0.1}{~GeV}$. This means that  below $\SI{1.4}{~GeV}$, the prescription is essentially provided by LEM, whereas above $\SI{1.6}{~GeV}$, the strength stems predominantly from the ReChi model.

The Hybrid model has been utilized in several studies to predict lepton-induced SPP on the nucleon and nucleus. The latter studies employed the relativistic plane-wave impulse approximation (RPWIA) and the relativistic distorted-wave impulse approximation (RDWIA) as the nuclear framework~\cite{Gonzalez-Jimenez18,Nikolakopoulos18,Gonzalez-Jimenez19,Nikolakopoulos23,Garcia-Marcos24}. In both approaches, the initial (bound) nucleons are described as relativistic mean-field wave functions,~i.e.,~solutions of the Dirac equation with relativistic potentials.  In RPWIA, the final (knocked-out) nucleon is treated as a plane wave, while in RDWIA, it is represented by a distorted wave, accounting for elastic final-state interactions (FSIs). In both cases, the pion was treated as a plane wave.

The implementation of the Hybrid Model in \nuwro for neutrino-nucleon pion production was presented in Ref.~\cite{Niewczas21}.  In this work, we incorporate the model into  \nuwro's nuclear framework, which employs a local Fermi gas (or the factorized spectral function approach) along with an intranuclear cascade model for the propagation of hadrons.

\section{\nuwro event generator}\label{nuwro}

The \nuwro MC event generator covers neutrino energies ranging from approximately $\SI{100}{MeV}$ to $\SI{100}{GeV}$~\cite{Juszczak:2005zs}. Neutrino-nucleus interactions are typically modeled within the impulse approximation scheme, wherein the initial interaction occurs on a bound nucleon followed by re-interactions of resulting hadrons inside the nucleus. Inelastic initial interactions are categorized as either resonant production (RES) for invariant hadronic masses $W\leq \SI{1.6}{GeV}$  or deep inelastic scattering (DIS) for $W>\SI{1.6}{GeV}$~\cite{Nowak:2006sx}.

\nuwro is optimized for sub-GeV neutrino beams, with RES dominated by the $\Delta (1232)$ excitation and its subsequent decay leading to SPP:
\begin{align}
    \nu~\text{N}^{'}\rightarrow \Delta~\ell,\qquad \Delta\rightarrow \pi~\text{N},
\end{align}
where N$^{(')}$ and $\ell$ represent the final (initial) nucleon and the scattered lepton, respectively.
Motivated by   quark-hadron duality, \nuwro has implemented an explicit model for $\Delta$ excitation,
while the strength provided by heavier resonances is included in an effective manner~\cite{Sobczyk:2004va}. Additional inelastic channels, such as two-pion production, are modelled with \pythia. By ``\pythia'', we mean that the lepton inclusive cross section is evaluated using the Bodek-Yang approach~\cite{Bodek:2003wd}, and the final hadronic state is obtained using the hadronization routines  of \pythia. In the case of SPP, a transition region from RES to DIS is modeled to ensure
a smooth transition between both regimes.

To achieve  a quantitative comparison between the predictions of the SPP model and experimental data, accurate models for both the initial state and FSIs must be incorporated. The effects of FSIs alter the observed spectra compared to the original model predictions. For instance, processes like pion absorption or production can induce changes in the observed final-state topology.

\subsection{Description of the $\Delta (1232)$ resonance}

The \nuwro $\Delta$ excitation model is formulated in terms of form factors obtained through a simultaneous fit to both ANL and BNL SPP data~\cite{Graczyk}. This fitting procedure was conducted for the $\nu_\mu \text{p}\rightarrow \mu^- \text{p}\pi^+$ reaction, assuming negligible contribution from the non-resonance background. Consequently, in the \nuwro  neutrino $\nu_\mu \text{p}\rightarrow \mu^- \text{p} \pi^+$ channel, there is no non-resonant background. However, in the $\nu_\mu \text{n}\rightarrow \mu^- \text{n}\pi^+$ and $\nu_\mu \text{n}\rightarrow \mu^- \text{p}\pi^0$ channels (and in analogous channels for $\bar\nu_\mu$ scattering, related by isospin symmetry), a non-resonant background is added incoherently as a fraction of the \pythia contribution, its size guided by the experimental data.
In the paragraphs below,  we describe the merging of both models to account for  inelastic events in \nuwro.

\subsection{$\Delta$-\pythia SPP model}
When modeling SPP channels, \nuwro implements the following cross section formula~\cite{Sobczyk:2004va}:
\begin{align}
    \sigmaSPP & =\beta(W)\sigmaRES + \alpha(W) \sigmaPySPP,
    \label{res-dis}
\end{align}
where $\sigmaRES$
stands for the $\Delta$ contribution and $\sigmaPySPP$ represents the SPP component of the DIS cross section as modeled by \pythia.
$\sigmaPySPP$ is defined with single pion production functions $f_\text{SPP}(W)$ extracted from \pythia as probabilities to get SPP final state as an outcome of the hadronization:
\begin{align}
    \sigmaPySPP & =f_\text{SPP}(W)\cdot\sigma^\text{DIS},
\end{align}
with $\sigma^\text{DIS}$ obtained from the Bodek-Yang approach.
The blending of the two is controlled by the functions  $\alpha (W)$ and $\beta(W)$. In \nuwro, the default option is that $\beta(W) = 1-\alpha (W)$ (more general assumption is available as well), where $\alpha (W)$ is a linear function of $W$:
\begin{equation}
    \alpha(W) = \begin{cases}
        \frac{W-\wth}{\wmin - \wth} \alpha_0                         & W < \wmin,               \\
        \frac{W-\wmin+\alpha_0\left(\wmax - W\right)}{\wmax - \wmin} & \wmin \leq W \leq \wmax, \\
        1                                                            & \wmax<W.
    \end{cases}
    \label{alpha}
\end{equation}
$\wth=M+m_\pi$ is a threshold for pion production (with $M$ the nucleon mass), $\wmin$ and $\wmax$  define the transition region, and their default  values are $\wmin=1.3$~GeV and $\wmax=1.6$~GeV. The linear function $\alpha(W)$ has the properties: $\alpha(\wth)=0$, $\alpha(\wmin)=\alpha_0$, and
$\alpha(\wmax)=1$. The parameter $\alpha_0$ defines the size of the non-resonant background. Its value is separately selected for each SPP channel, as shown in Table~\ref{tab:sppval}.

\begin{table}[hbt!]
    \centering
    \begin{tabular}{@{}cc@{}}
        \toprule
        Channel                               & $\alpha_0$           \\ \hline
        $\nu p\rightarrow \ell^-p\pi^+$       & \multirow{2}{*}{0.0} \\
        $\bar{\nu} n\rightarrow \ell^+n\pi^-$ &                      \\  \hline
        $\nu n\rightarrow \ell^-n\pi^+$       & \multirow{2}{*}{0.2} \\
        $\bar{\nu} p\rightarrow \ell^+p\pi^-$ &                      \\ \hline
        $\nu n\rightarrow \ell^-p\pi^0$       & \multirow{2}{*}{0.3} \\
        $\bar{\nu} p\rightarrow \ell^+n\pi^0$ &                      \\ \hline
    \end{tabular}
    \caption{\nuwro default values of the $\alpha_0$ parameter (see Eq.~\ref{alpha}).
    }\label{tab:sppval}
\end{table}

\subsection{\nuwro algorithm}

In the RES region $W<\wmax$ \nuwro combines SPP model with other inelastic channels, such as two-pion production, extracted from \pythia. The generation of inelastic events is achieved through the following steps:

\begin{enumerate}
    \item Random phase space sampling : The algorithm begins by randomly selecting a point within the two-dimensional available kinematic phase space $(W, Q^2)$
          for a given neutrino energy, with the constraint $W<\wmax$.
          This sampled point defines the $W$ value for the resulting hadronic system.

    \item Hadronization and event weight : At the selected phase space point, \pythia is invoked for hadronization, where the hadronic energy and momentum are converted into final-state particles.
          \begin{enumerate}
              \item Non-SPP Event: If the final state is not SPP,
                    the event is accepted
                    and assigned with
                    a weight equal to
                    $\frac{\displaystyle \diff^2\sigma^\text{DIS}}{\displaystyle \diff Q^2\diff W}$ multiplied by the available phase space in $(W, Q^2)$.
              \item SPP Event: If the final state
                    generated by \pythia is SPP, the event is assigned with a weight
                    $\frac{1}{f_\text{SPP}(W)} \frac{\diff^2\sigmaSPP}{\diff Q^2\diff W}$
                    multiplied by the available phase space.

                    The blending defined in
                    Eq.~\ref{res-dis}
                    is realized by probabilistically choosing an either $\Delta$ or \pythia origin of the event according to $\beta(W):\alpha(W)$ ratio.
                    If the former is
                    chosen, the kinematics of the final hadronic state is replaced by the outcome of $\Delta$ decay, incorporating information regarding angular correlations. If the latter is selected, the event is stored as it is.

          \end{enumerate}
\end{enumerate}

\subsection{Hybrid model in \nuwro}\label{sec:hynuwro}

\begin{figure}[!tb]
    \centering
    \includegraphics[width=\figwid\textwidth]{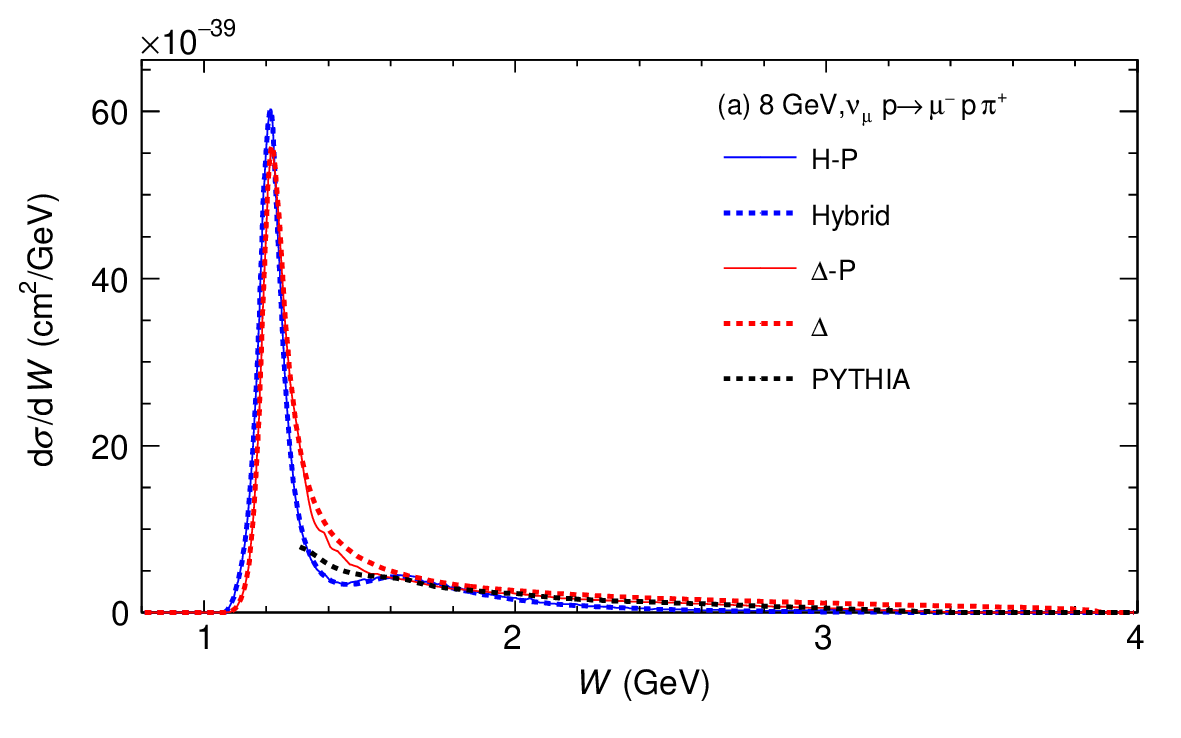}
    \includegraphics[width=\figwid\textwidth]{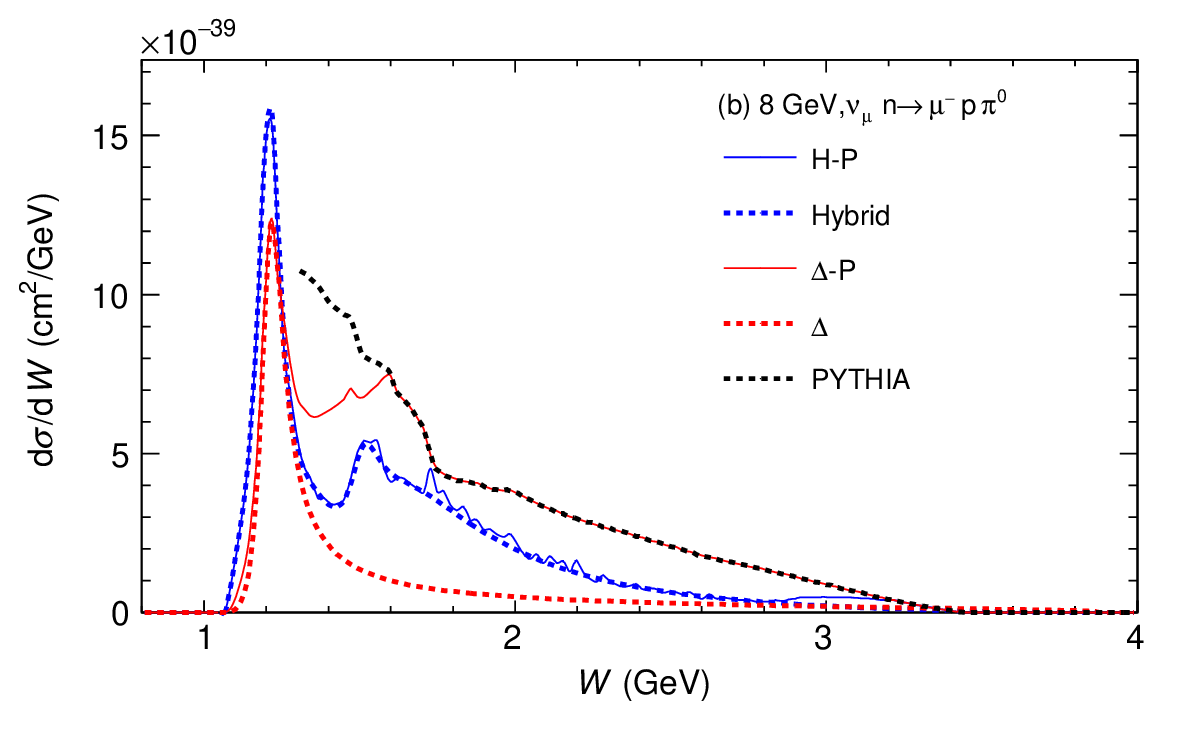}
    \includegraphics[width=\figwid\textwidth]{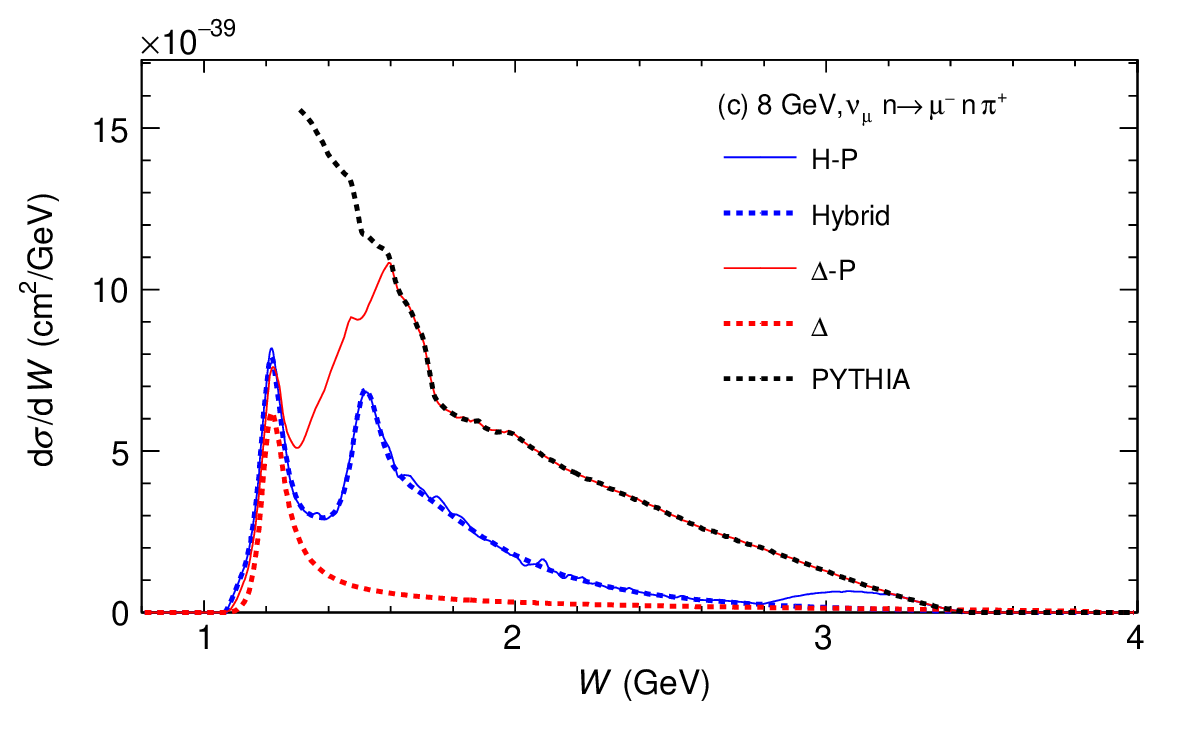}
    \caption{Predictions for SPP by an 8~GeV neutrino off the nucleon for the models discussed in this paper for three possible channels: (a) $\nu_\mu\textrm{p}\rightarrow\mu^-\textrm{p}\pi^+$, (b) $\nu_\mu\textrm{n}\rightarrow\mu^-\textrm{p}\pi^0$, and (c) $\nu_\mu\textrm{n}\rightarrow\mu^-\textrm{n}\pi^+$. The transition regions in $W$ are 1.3--1.6 $\si{GeV}$ and  2.8--3.2  $\si{GeV}$ for the $\Delta$-P and
        H-P models, respectively. The wiggle in $\Delta$-P and H-P curves is due to numerical  fluctuations in evaluating $f_\text{SPP}(W)$. Note the cross sections at the $\Delta$ peaks predicted by the $\Delta$ model: $9:2:1$, as expected from isospin arguments~\cite{Adler:1968tw, Barish:1975bw, Rein:1980wg}.
    }
    \label{fig:trans}
\end{figure}

Considering the Hybrid model's ability to describe SPP across a wide kinematic range, an expansion of \nuwro's current transition region to higher values of $W$ appears reasonable. A motivation to do this follows from what is seen in Fig.~\ref{fig:trans}, where
we compare the predicted SPP cross sections ($\dv{\sigma}{W}$) using different models (another set of predictions with lower neutrino energy 3~GeV is shown in Fig. \ref{fig:3GeVtrans} in Appendix~\ref{sec:appSPP}). Around $W = \SI{1.5}{~GeV}$, the Hybrid model exhibits a second peak, particularly noticeable in the p$\pi^0$ and n$\pi^+$ channels (Figs.~\ref{fig:trans}b and~\ref{fig:trans}c), attributed to contributions from the second family of nucleon resonances, $D_{13}(1520)$, and $S_{11}(1535)$.
The $\Delta$ model does not incorporate these contributions. In the results of the $\Delta$-P model, a structure can be observed that might mistakenly appear to arise from the second resonance region. However, this structure is merely an artifact of the transition region and holds no physical significance, as discussed in Ref.~\cite{Sobczyk:2004va}.
Moreover, the $\Delta$ model requires an earlier transition to the \pythia model to include the cross section strength in the second resonance region and beyond, which is not accounted for in the pure $\Delta$ model.

To minimize or eliminate Pythia contributions to the SPP channel, the transition region would ideally be placed at very high $W$ values. However, in the current NuWro algorithm governing single pion production and DIS events, such a high $W$ transition
makes the code significantly more time consuming. A remedy would be to replace the current RES/DIS transition defined according to values of $W$, by SPP/DIS transition defined by topology of final states, but this requires a change of the basic structure of the code. In this study of the Hybrid model, we set $\wmin = 2.8$ GeV and $\wmax = 3.2$ GeV. Additionally, as the Hybrid model already contains contributions from the non-resonant background, in its implementation in NuWro, we set $\alpha_0=0$ (see Eq.~\ref{alpha}). Henceforth, we will refer to the \nuwro $\Delta$ and Hybrid models incorporating the transition to \pythia as $\Delta$-\pythia and Hybrid-\pythia, respectively.

\subsection{Nuclear modeling}
The nuclear modeling in \nuwro assumes quasi-free scattering, including the target nucleon binding energy and Fermi motion, both modeled in a similar fashion as in quasielastic (QE) scattering. \nuwro offers various options for describing bound nucleons. In this study, benchmark computations will be conducted using the local Fermi-Gas (LFG) and effective spectral function (ESF) models~\cite{Ankowski:2005wi}.
Currently, \nuwro cannot fully consider effects related to $\Delta$ self-energy in nuclear matter~\cite{OSET1987631}; the $\Delta$ width is assumed to be the same as in vacuum. However, some of these effects are considered in the nuclear cascade model (see text below).

In the ESF approach, the nucleon momentum is sampled from a probability density distribution identical to that defined in the hole spectral function (SF) approach~\cite{Benhar:1989aw}. The momentum-dependent binding energy is evaluated as an average provided by the SF.
The LFG approach evaluates the nucleon momentum based on local density information at the interaction point. The target nucleon is assumed to be in a potential equal to $E_\text{F}+V$, where $E_\text{F}$ is the Fermi energy and $V=8$~MeV for carbon.

\begin{figure}[!tb]
    \centering
    \includegraphics[width=\figwid\textwidth]{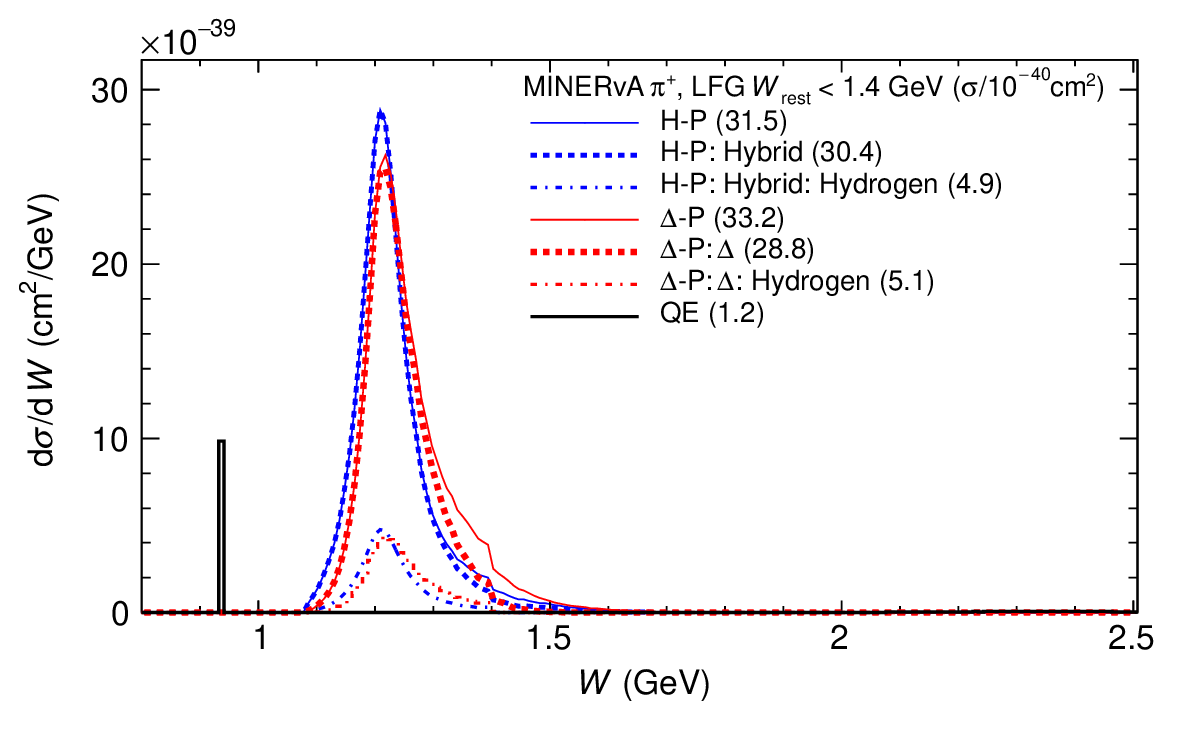}
    \caption{Predictions for the $W$ distribution in the MINERvA $\pi^+$ measurements~\cite{MINERvA:2014ogb} from the models discussed in this paper.  The notation ``A:~B'' refers to the contribution of B in A. Beside the $\Delta$ and Hybrid models, the remaining contributions include single- and multi-pion  production from \pythia. The SPP on the hydrogen target component is also shown. The initial state is modelled by LFG except for the QE component which is by SF. The predictions are restricted to $W_\textrm{rest}<1.4~\textrm{GeV}$, where $W_\textrm{rest}$ is the $W$ calculated assuming an initial nucleon at rest.
        The integrated cross section for each component is provided in parentheses, with units of $\SI{e-40}{cm^2}$. This convention also applies to Fig.~\ref{fig:mbW}, Fig.~\ref{fig:TKIW} and Fig.~\ref{fig:T2K_TKI_W}.}\label{fig:mpiW}
\end{figure}

When comparing to the experimental SPP data, it is essential to consider the effects of FSI. FSI is modeled in \nuwro using an intranuclear cascade model~\cite{Golan:2012wx,Niewczas:2019fro}.
The cornerstone of this cascade model lies in the microscopic hadron-nucleon cross sections. Specifically for pions, these cross sections, which are density-dependent, are derived from the Oset-Salcedo model~\cite{Salcedo:1987md}. Pions traverse through the nucleus in steps of $0.2$ fm. At each step, the probability of a pion-nucleon interaction is calculated, and an MC algorithm determines if and how the interaction occurs (absorption, charge exchange, or elastic scattering).

\section{Comparison with data}\label{datacomparison}

\begin{figure}[!tb]
    \centering
    \includegraphics[width=\figwid\textwidth]{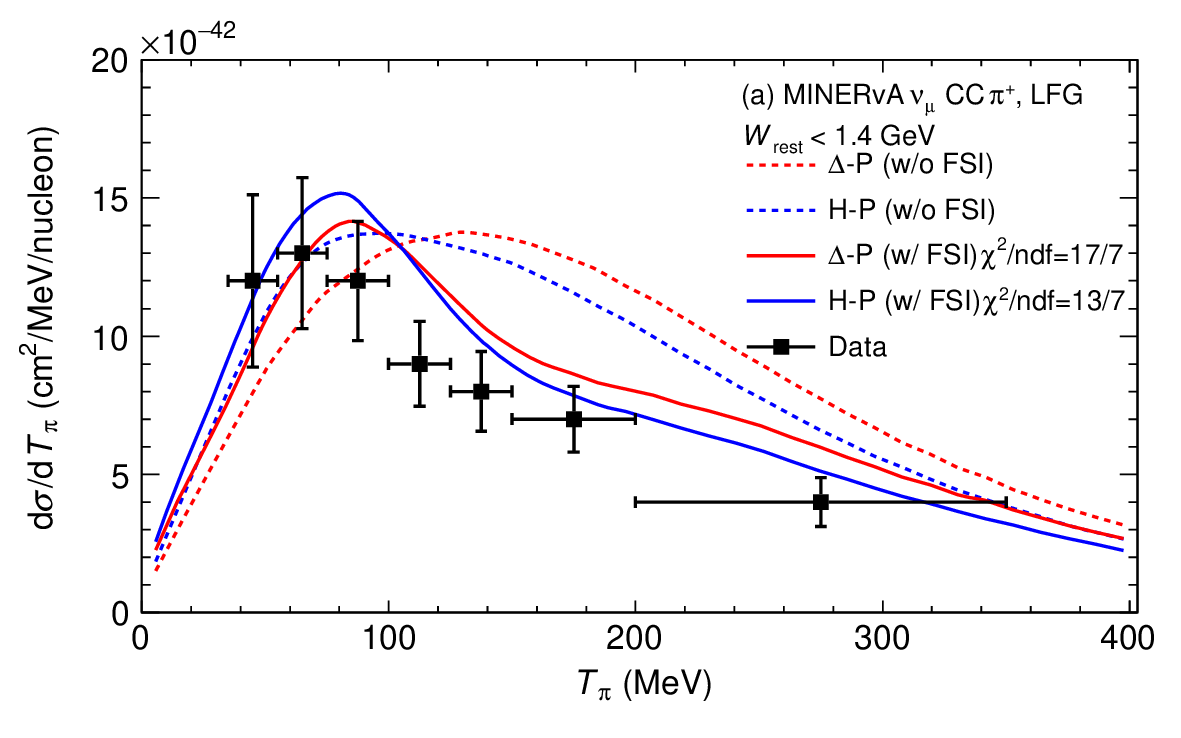}
    \includegraphics[width=\figwid\textwidth]{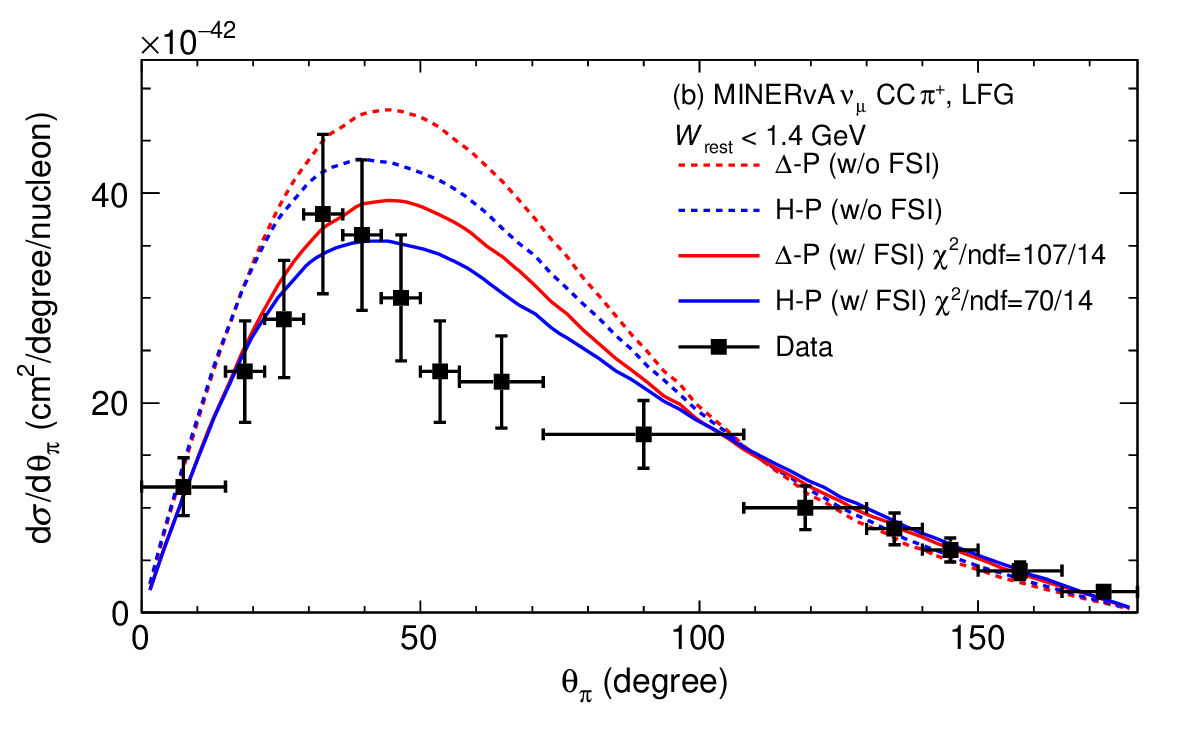}
    \caption{Predictions of the $\Delta$-P and H-P models, with and without FSI, compared to the MINERvA measurement~\cite{MINERvA:2014ogb} for the $\pi^+$ (a) kinetic energy, $T_\pi$,  and (b) polar angle, $\theta_\pi$, distributions. Both the measurements and predictions are restricted to $W_\textrm{rest}<1.4~\textrm{GeV}$ (cf. Fig.~\ref{fig:mpiW}). }
    \label{fig:minervapip}
\end{figure}

\begin{figure}[!tb]
    \centering
    \includegraphics[width=\figwid\textwidth]{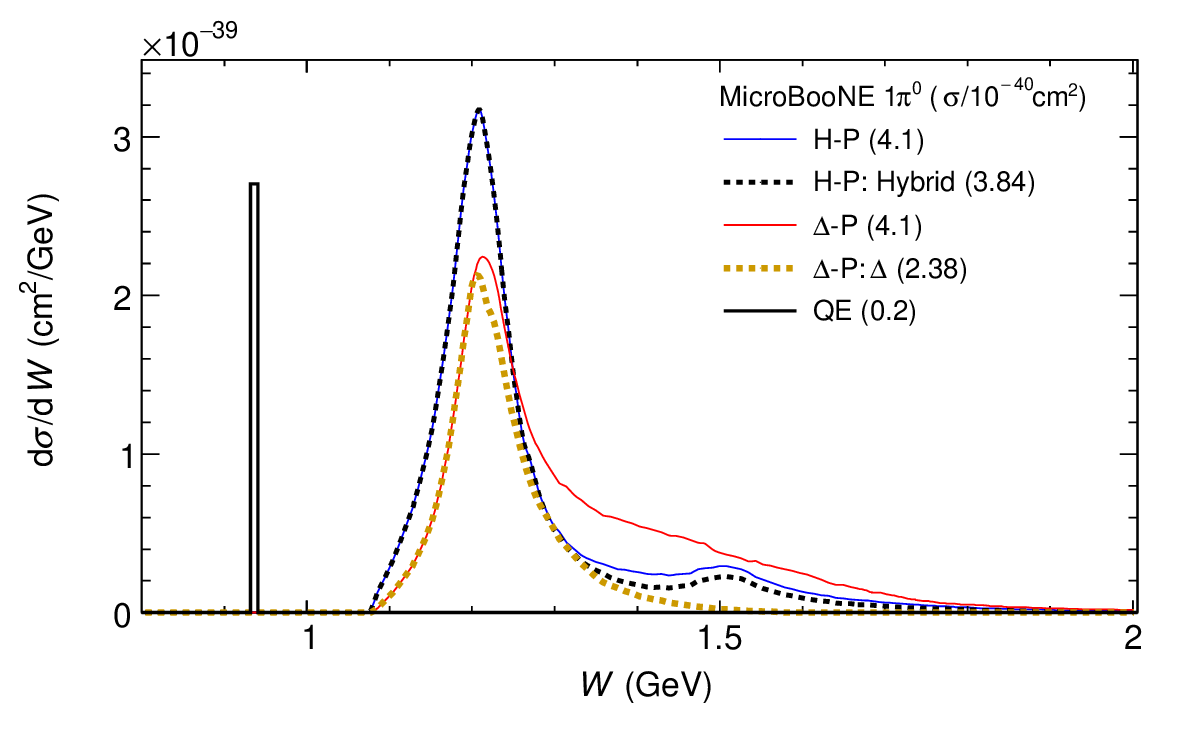}
    \caption{Predictions for the $W$ distributions in the MicroBooNE CC$\pi^0$ measurements from the models discussed in this paper.}
    \label{fig:mbW}
\end{figure}

\begin{figure}[!tb]
    \centering
    \includegraphics[width=\figwid\textwidth]{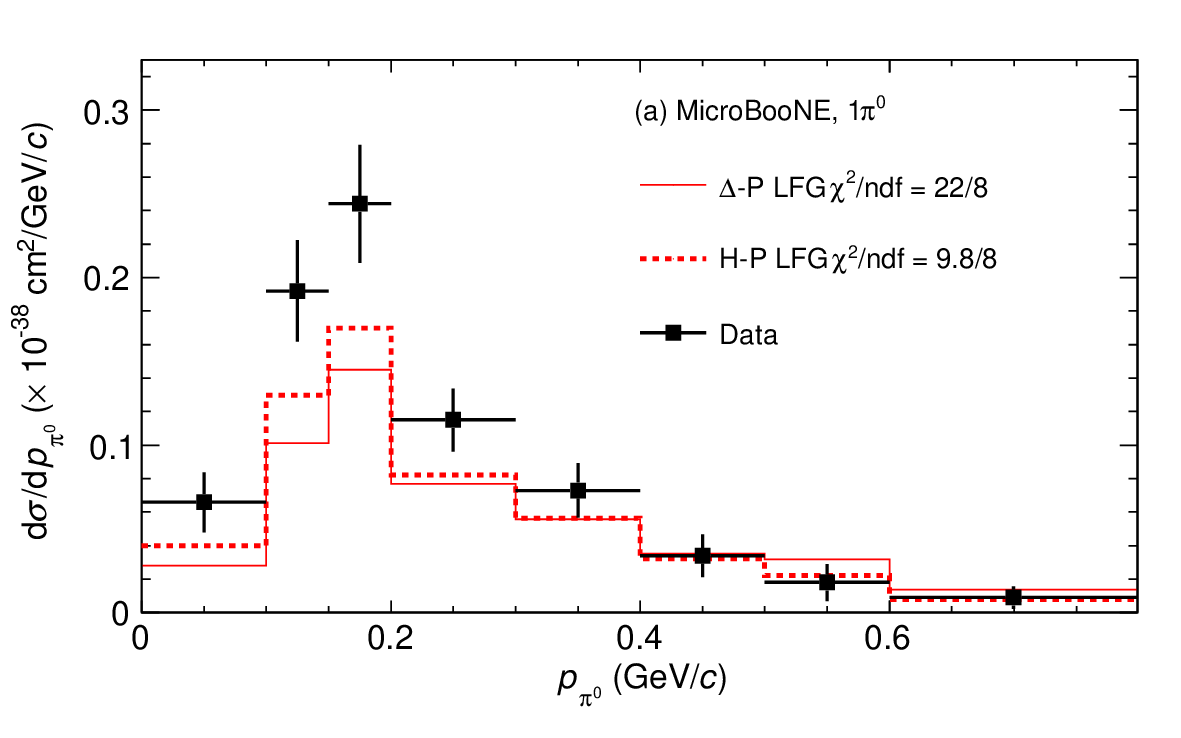}
    \includegraphics[width=\figwid\textwidth]{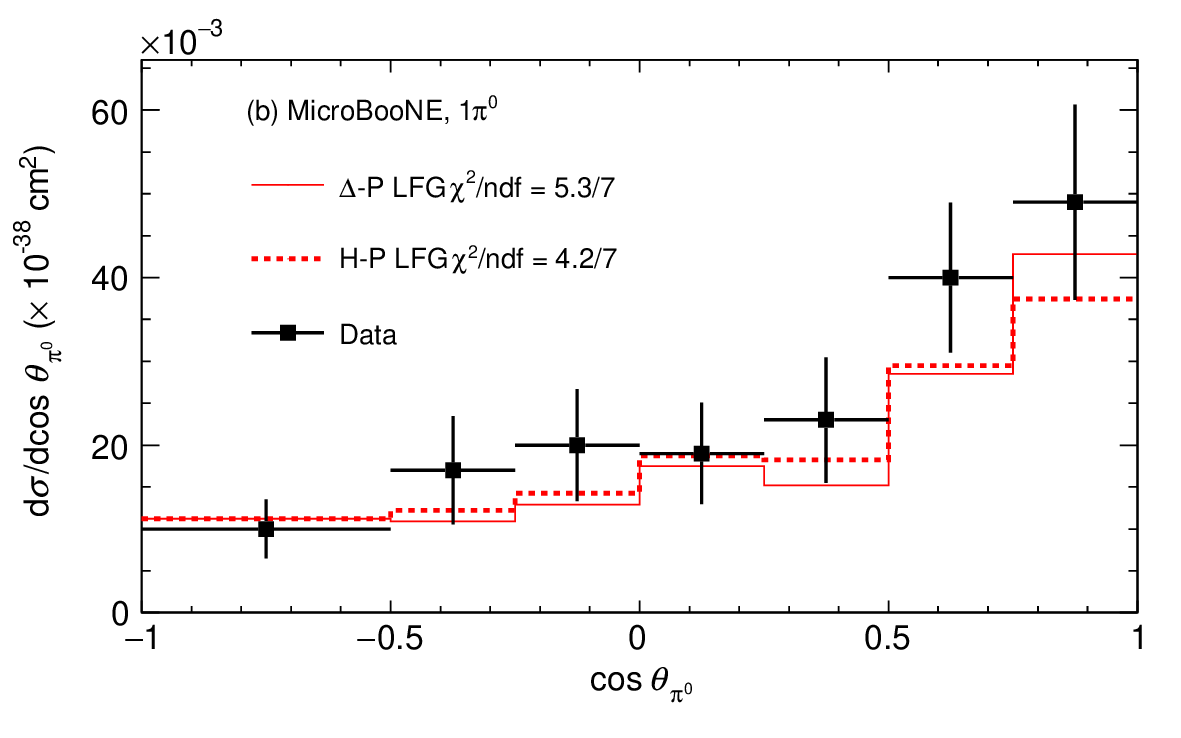}
    \caption{Predictions of the $\Delta$-P and H-P models compared to the MicroBooNE measurement~\cite{microboone:pi0} for the $\pi^0$ (a) momentum, $p_\pi$,  and (b) cosine polar angle, $\cos\theta_\pi$, distributions.}
    \label{fig:mbpi0}
\end{figure}

\subsection{Pion kinematics}

Using a neutrino beam with an average energy of $\SI{3}{GeV}$, the MINERvA experiment systematically measured charged-current (CC) pion production from a hydrocarbon target (cf. Ref.~\cite{MINERvA:2021csy} for a review). For the CC semi-inclusive single $\pi^+$ measurement~\cite{MINERvA:2014ogb}, the predicted $W$ distributions by the $\Delta$-P and H-P models are shown in Fig.~\ref{fig:mpiW} (QE events contribute to the measurement through pion production during FSI. For a comprehensive  data-model comparison, this contribution is included in the figure), and the pion kinematics are compared to data in Fig.~\ref{fig:minervapip}.
This channel is dominated by the $\Delta^{++}$ decay shown in Fig.~\ref{fig:trans}a.
As the measurement is limited to $W_\textrm{rest}<1.4~\textrm{GeV}$, where $W_\textrm{rest}$ is the $W$ defined via lepton kinematics assuming the initial nucleon at rest,
contributions from mechanisms beyond the $\Delta$ region are largely suppressed.  Figure~\ref{fig:mpiW} suggests that the results obtained from the models discussed in this study are expected to show minimal differences.
In Fig.~\ref{fig:minervapip}, in addition to the usual improvement seen by switching on FSI~\cite{Gonzalez-Jimenez18,Nikolakopoulos18}, which essentially consists in a redistribution of the strength towards lower bins of the pion energy,
we see that the H-P model exhibits an improvement in both the pion energy and angular distributions compared to the $\Delta$-P model.

The H-P model performs significantly better compared to the $\Delta$-P model against the MicroBooNE CC$\pi^0$ production data~\cite{microboone:pi0}, which is predominantly driven by $\Delta^{+}$.
A similar comparison between the $\Delta$-P  and H-P models (cf. Fig.~\ref{fig:trans}b)  is shown in Fig.~\ref{fig:mbW} for MicroBooNE. The enhanced $\Delta$ strength from the H-P model leads to a better agreement with the MicroBooNE data (Fig.~\ref{fig:mbpi0}).

Note that, despite these improvements, the \nuwro model does not fully reproduce data; for a discussion see also Ref.~\cite{Sobczyk:2014xza}.

\subsection{Transverse kinematics imbalance}

In addition to pion kinematics, the MINERvA measurement of the transverse kinematic imbalance (TKI) in neutral-pion production has highlighted significant challenges, suggesting issues at the pion production level~\cite{MINERvA:2020anu}.
TKI is a methodology based on momentum conservation considerations. It involves assessing the disparity between the observed transverse momentum of final-state particles and what would be expected from neutrino interactions with free nucleons.
This kinematic mismatch~\cite{Lu:2015hea, Lu:2015tcr}, along with its longitudinal and three-dimensional variations~\cite{Furmanski:2016wqo, Lu:2019nmf}, and the derived asymmetries~\cite{Cai:2019jzk}, have contributed to extracting valuable information about the particles involved in the interaction and the underlying nuclear processes. Unlike the recent pionless measurements by T2K~\cite{T2K:2018rnz} and MINERvA~\cite{MINERvA:2018hba, MINERvA:2019ope}, the $\pi^0$ and $\pi^+$ production by MINERvA~\cite{MINERvA:2020anu} and T2K~\cite{T2K:2021naz} have shown significant model deficiency in the kinematic region populated by events devoid of FSI (see discussions below).

\begin{figure}[tb]
    \centering
    \includegraphics[width=\figwid\textwidth]{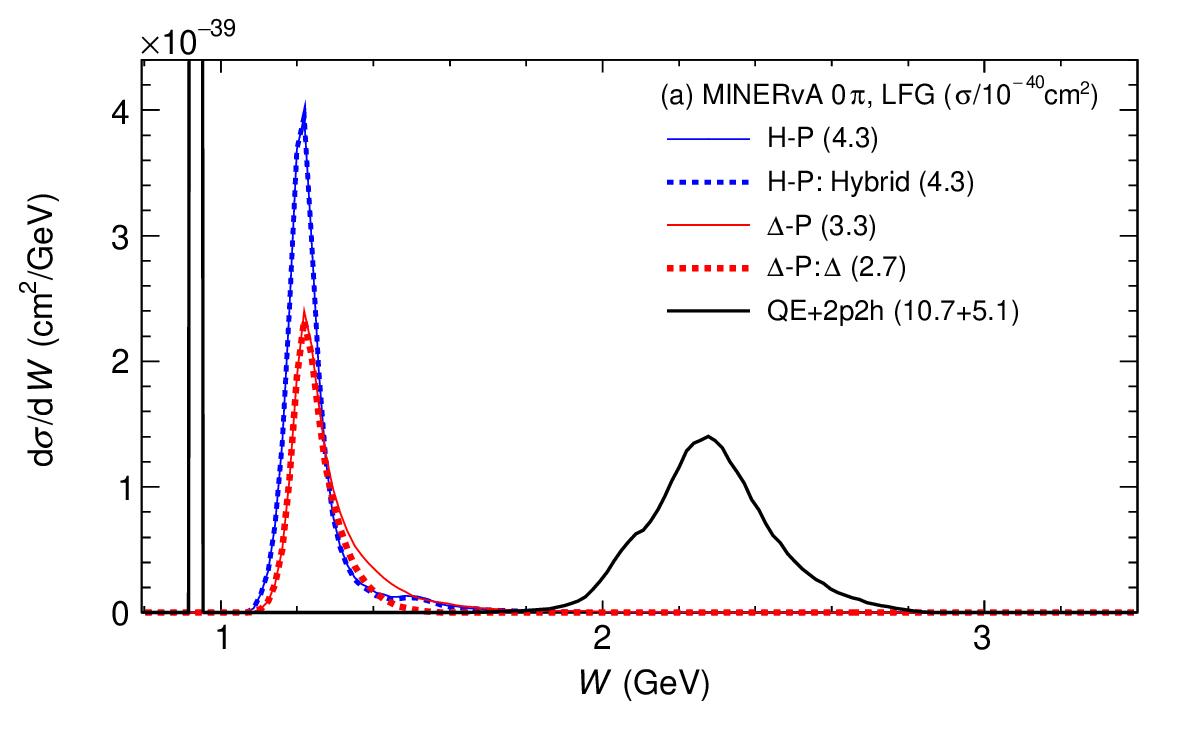}
    \includegraphics[width=\figwid\textwidth]{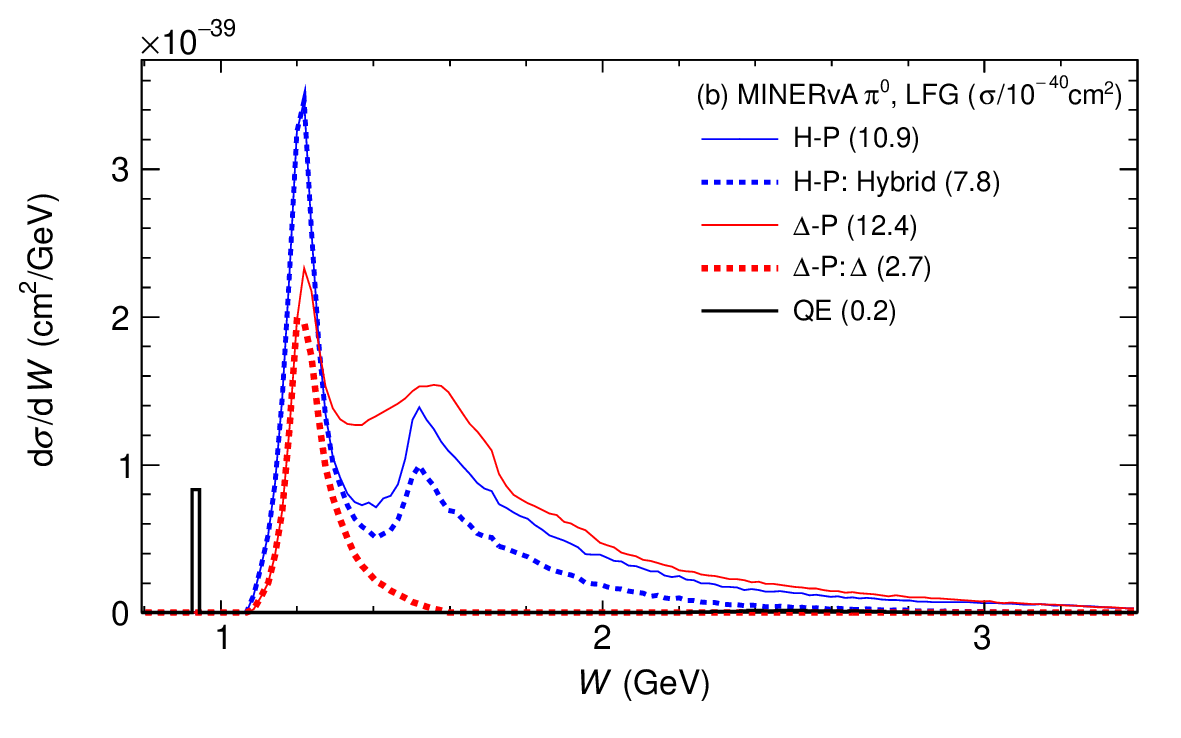}
    \caption{Similar to Fig.~\ref{fig:mpiW} but for the MINERvA (a) $0\pi$~\cite{MINERvA:2018hba} and (b) $\pi^0$~\cite{MINERvA:2020anu} TKI  signals.
    }
    \label{fig:TKIW}
\end{figure}

Consider a neutrino-nucleus interaction:
\begin{equation}
    \nu + \textrm{A} \rightarrow \ell + \textrm{N} + \textrm{X},
\end{equation}
where A and X are the initial nucleus and final nuclear remnant, respectively, $\ell$ is the CC lepton, and $\textrm{N}$ refers to a proton in the CC pionless channel or the $p+\pi$ system in a CC pion production.
The transverse boosting angle, $\delta \alpha_\textrm{T}$~\cite{Lu:2015tcr}, and the emulated (initial state) nucleon momentum, $p_\textrm{N}$~\cite{Furmanski:2016wqo, Lu:2019nmf}, are defined as:
\begin{gather}
    \delta \vec{p}_\textrm{T} = \vec{p}^{\,\ell}_\textrm{T} + \vec{p}^{\,\textrm{N}}_\textrm{T} \label{dpt}, \\
    \delta \alpha_\textrm{T} = \arccos\frac{ - \vec{p}^{\,\ell}_\textrm{T} \cdot \delta \vec{p}_\textrm{T}}{\left|\vec{p}^{\,\ell}_\textrm{T}\right| \left|\delta \vec{p}_\textrm{T}\right|} \label{dat}, \\
    \delta p_\textrm{L} = \frac{R^2 -\delta \vec{p}_\textrm{T}^{\,2} - M_\textrm{A-1}^{*2} }{2R} \label{dpl}, \text{with: } \\
    R \equiv M_\textrm{A} + p^\ell_\textrm{L} + p^\textrm{N}_\textrm{L} - E^\ell - E^\textrm{N} \label{Rdef} , \\
    p_\textrm{N} = \sqrt{\delta \vec{p}_\textrm{T}^{\,2} +  \delta p_\textrm{L}^2} \label{pn} .
\end{gather}
In these definitions, $\vec{p}^{\,\kappa}_{\textrm{T}}$($p^\kappa_\textrm{L}$) represents the transverse (longitudinal) component of the particle $\kappa$ in the final state relative to the direction of the neutrino. $\delta \vec{p}_\textrm{T}$ is the transverse component of the missing momentum between the initial state and final state. Its longitudinal counterpart, $\delta p_\textrm{L}$, is dependent on the initial nucleus mass $M_\textrm{A}$, as well as the energies of the lepton and hadron, $E^\ell$ and $E^\textrm{N}$, along with the mass of the resulting nuclear remnant, $M_\textrm{A-1}^{*}$, given by:
\begin{equation}
    M_\textrm{A-1}^{*} = M_\textrm{A} - M_\textrm{n} + b.
\end{equation}
Here, $M_\textrm{n}$ is the neutron mass and $b = \SI{28.7}{MeV}$~\cite{Furmanski:2016wqo, Lu:2019nmf} represents the average excitation energy.

\begin{figure}[tb]
    \centering
    \includegraphics[width=\twofigwid\textwidth]{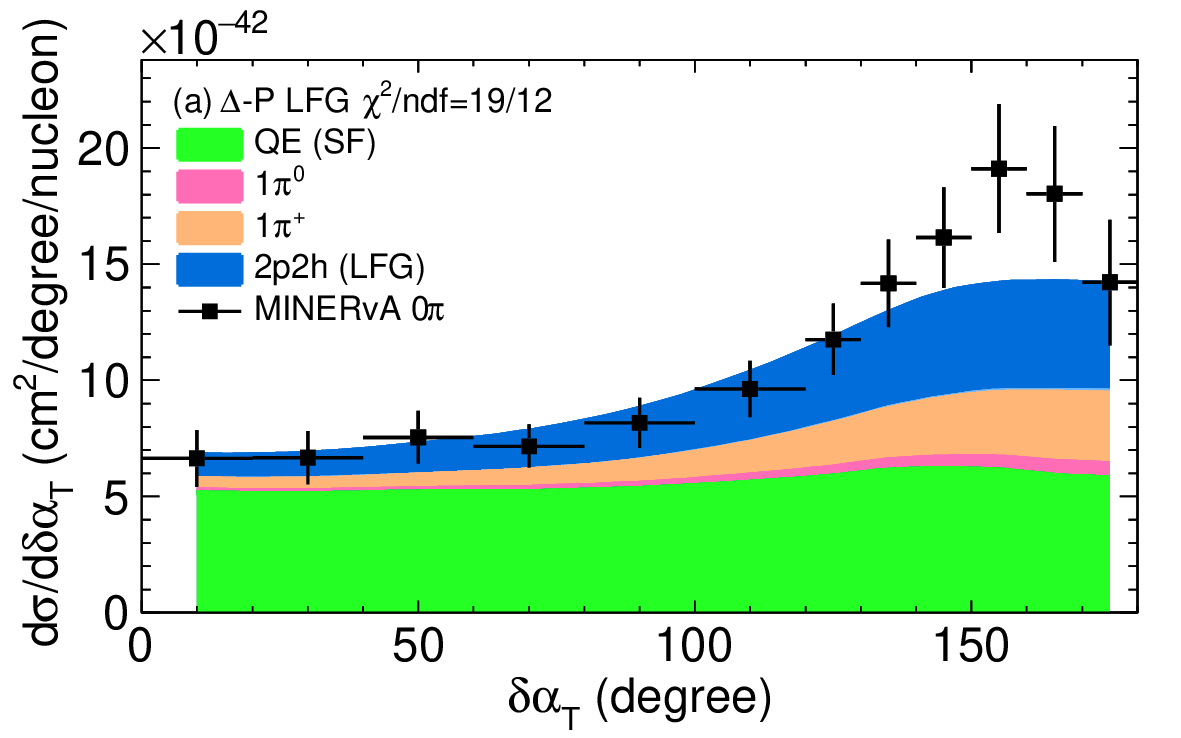}
    \includegraphics[width=\twofigwid\textwidth]{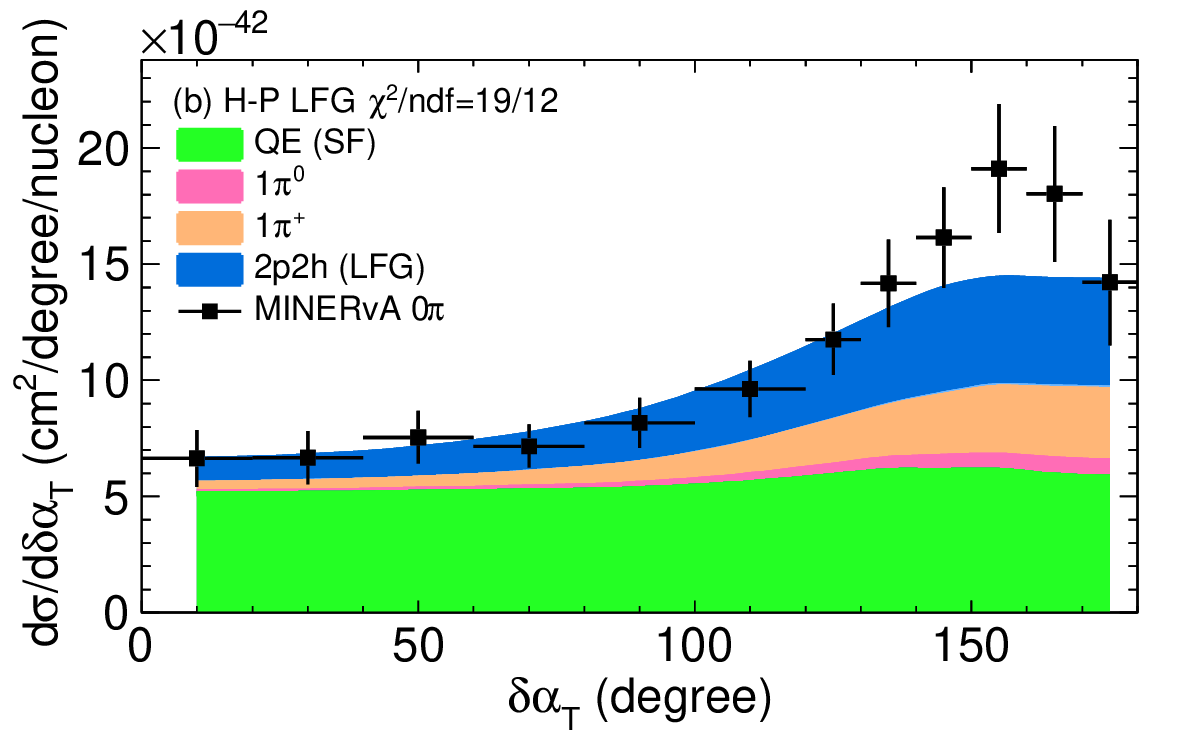}

    \includegraphics[width=\twofigwid\textwidth]{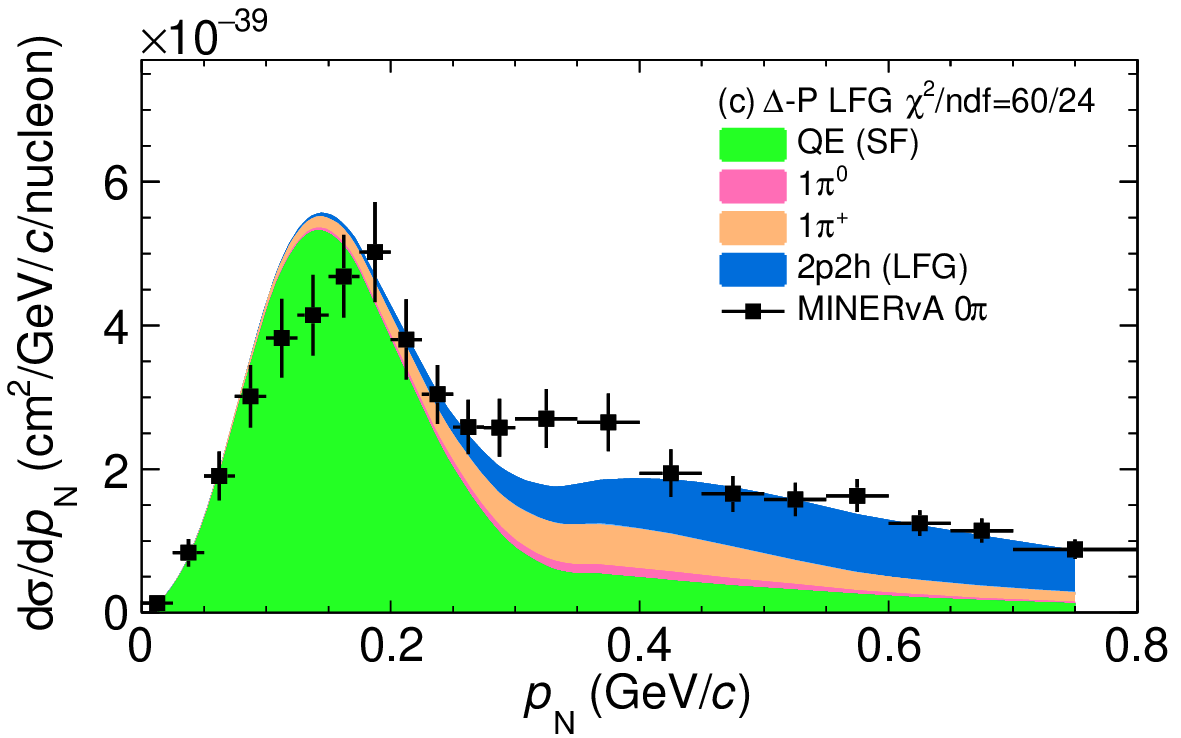}
    \includegraphics[width=\twofigwid\textwidth]{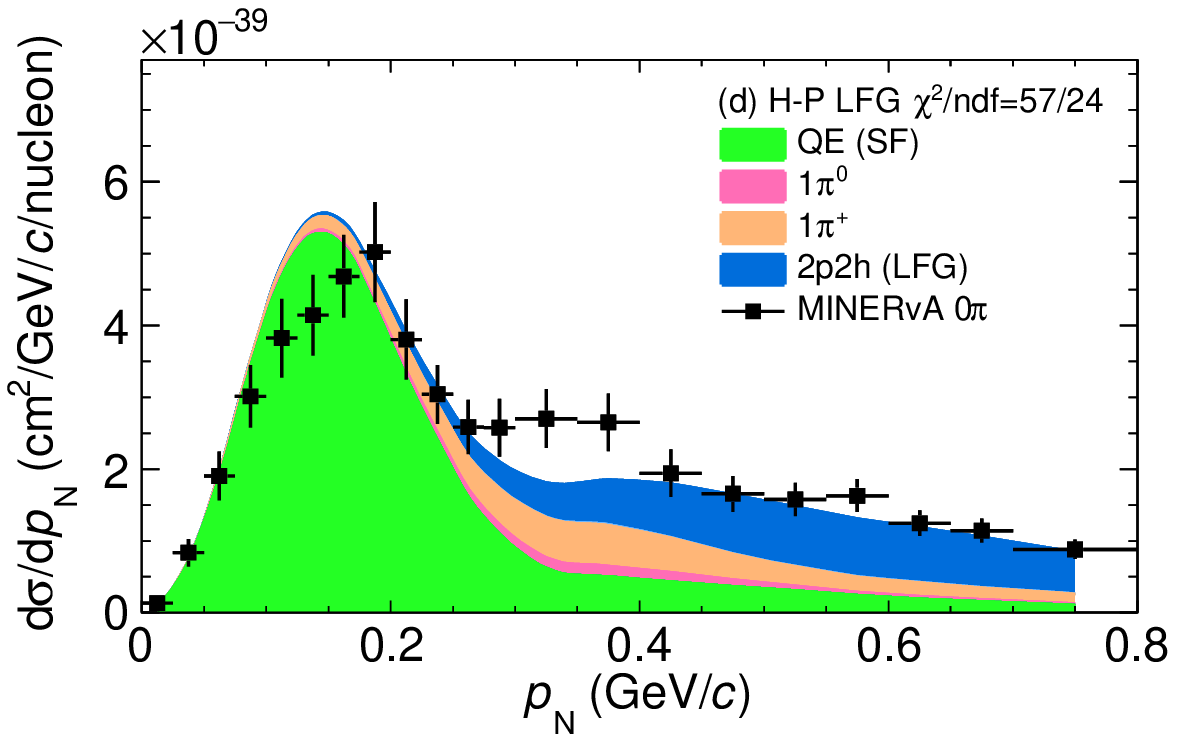}

    \includegraphics[width=\twofigwid\textwidth]{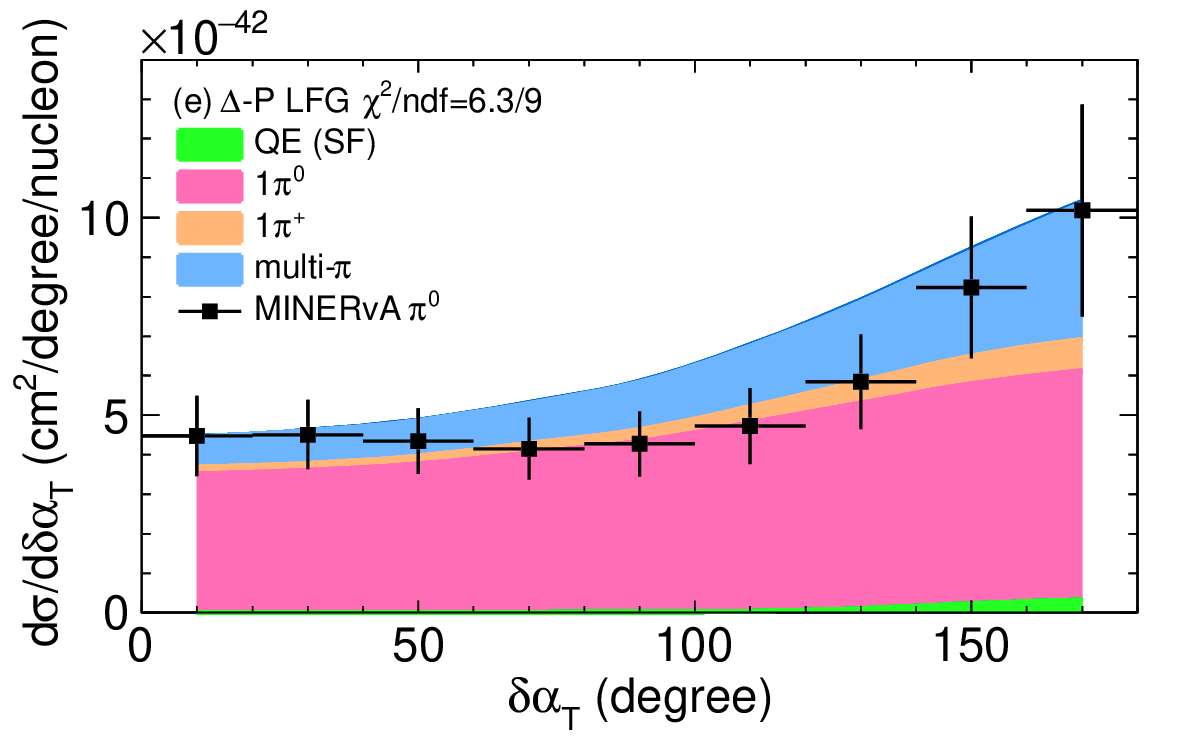}
    \includegraphics[width=\twofigwid\textwidth]{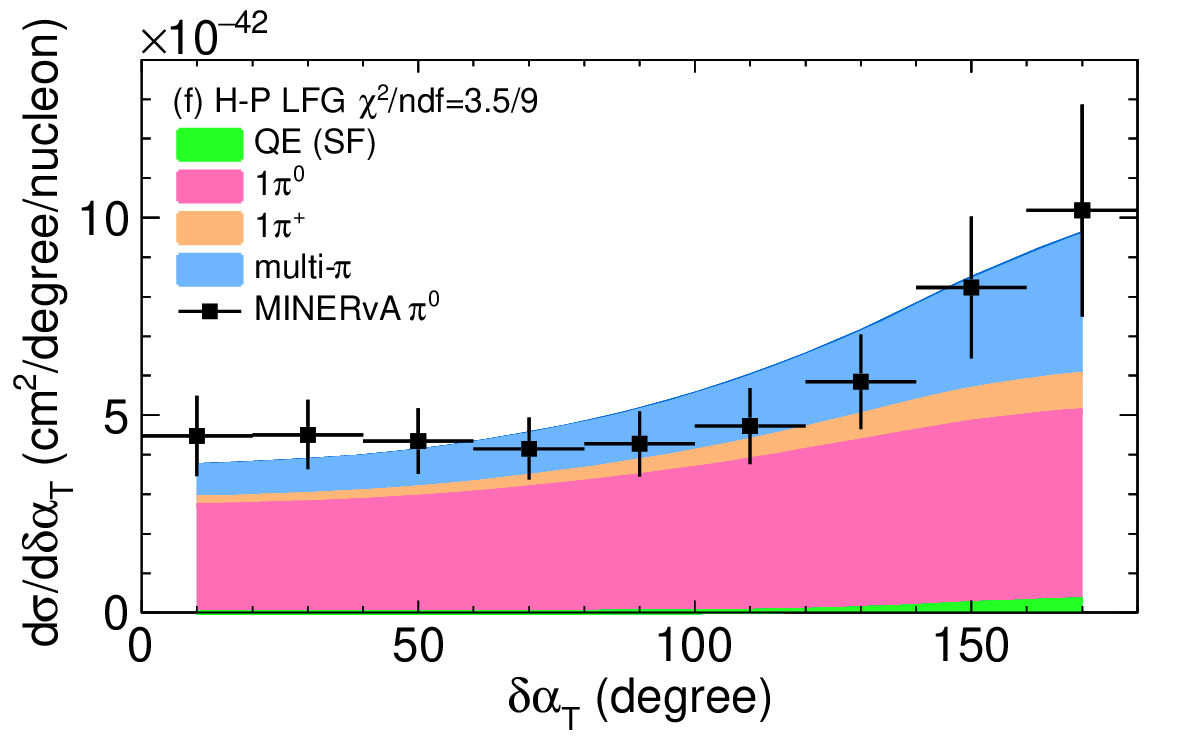}

    \includegraphics[width=\twofigwid\textwidth]{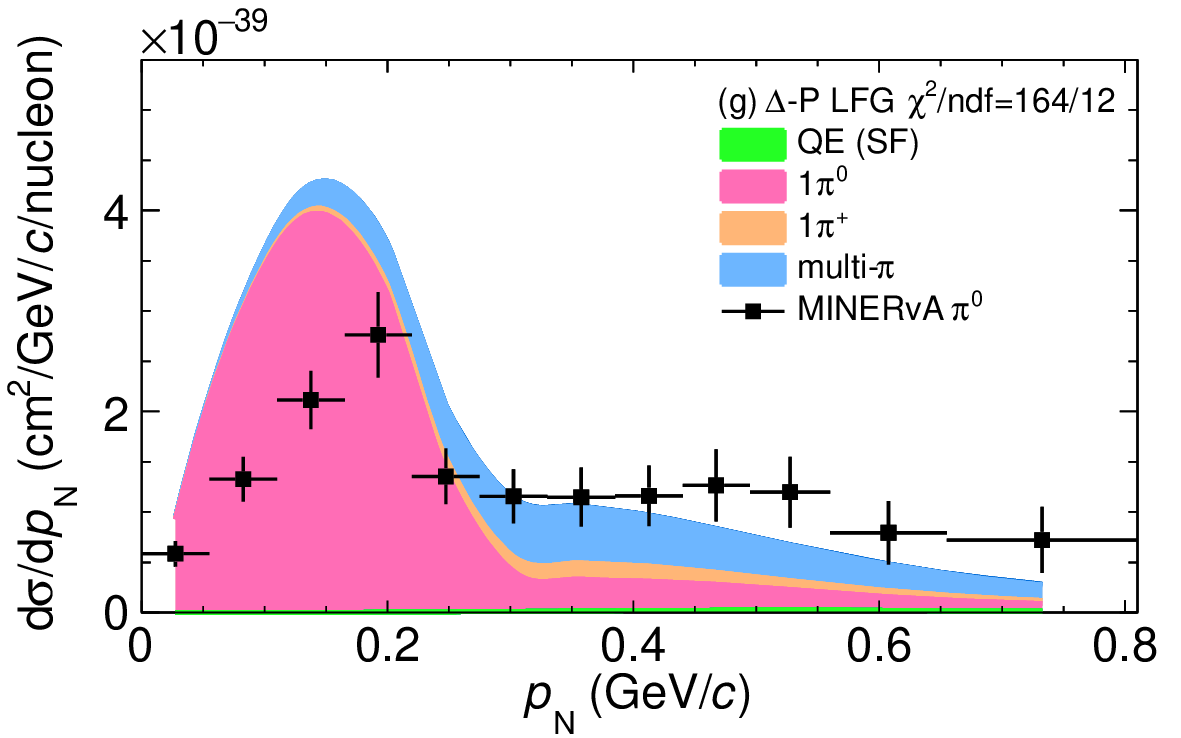}
    \includegraphics[width=\twofigwid\textwidth]{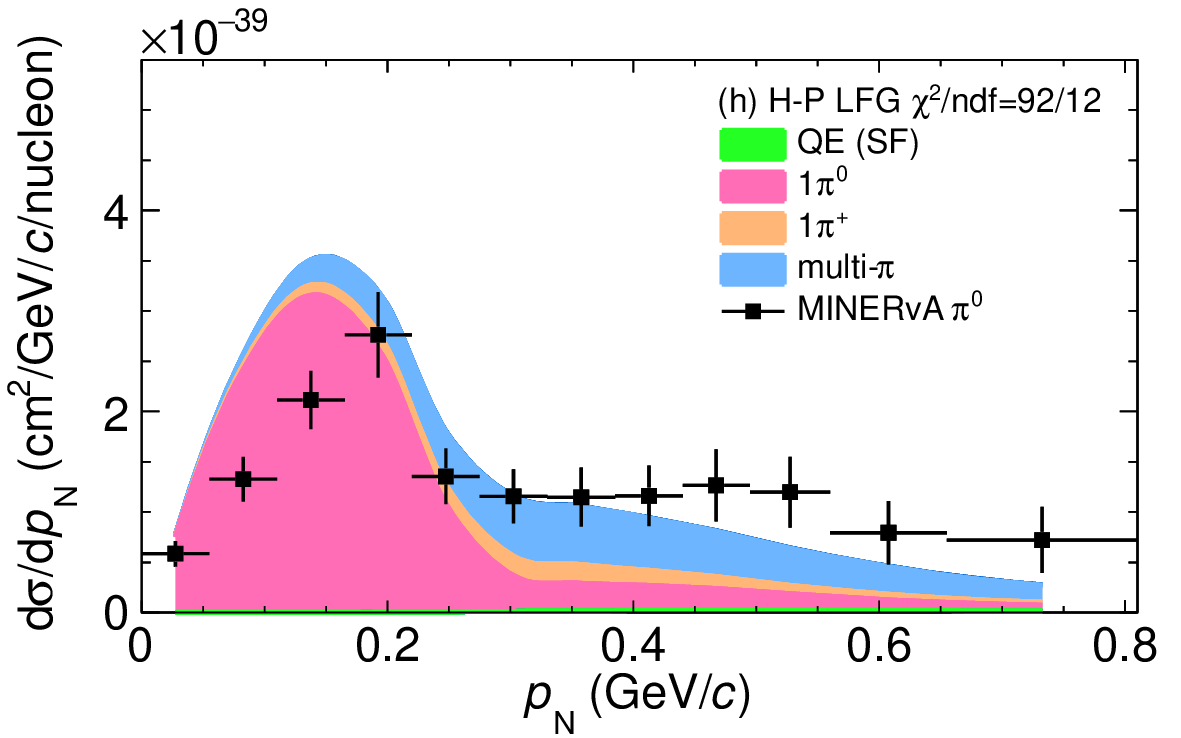}

    \caption{$\Delta$-P (\textit{left}) and H-P (\textit{right}) model predictions for the $\delta\alpha_\textrm{T}$ and $p_\textrm{N}$ distributions in the MINERVA  0$\pi$~\cite{MINERvA:2018hba} (\textit{upper}) and  $\pi^0$~\cite{MINERvA:2020anu}  (\textit{lower}) measurements. The prediction is decomposed according to the interaction type (QE, 2p2h) and topology (1 $\pi$, multi-$\pi$) prior to FSI. }
    \label{fig:NuWro_LFG_TKI}
\end{figure}

For one-body currents, $\delta p_\textrm{T}$~\cite{Lu:2015tcr} probes the transverse projection of the Fermi motion of the struck nucleon in the absence of FSIs, with its angle, $\delta \alpha_\textrm{T}$, mostly uniformly distributed (except for centre-of-mass effects) due to the isotropy of Fermi motion.
Deviations from a uniform $\delta \alpha_\textrm{T}$ distribution may indicate the influence of FSIs and potential contributions from two-body currents.
$p_\textrm{N}$ offers the fine details of the Fermi motion; the change from $\delta p_\textrm{T}$ to $p_\textrm{N}$ can be considered as a correction of the order of $\mathcal{O}(20\%)$~\cite{Yang:2023dxk}.

In the following, we focus on comparing our results with the MINERvA $\pi^0$ TKI measurement~\cite{MINERvA:2020anu}, while using the $0\pi$ measurement~\cite{MINERvA:2018hba, MINERvA:2019ope} as a control group. Further comparison with the T2K $\pi^+$ measurement~\cite{T2K:2021naz} is available in Appendix~\ref{sec:appT2K}. The respective signal definitions are:
\begin{itemize}
    \item For CC-$\pi^0$~\cite{MINERvA:2020anu}:
          \begin{enumerate}
              \item $\nu_\mu + \textrm{A} \rightarrow \mu^- + \text{p} + \pi^0 + \textrm{X}$, requiring one $\mu$, and at least one $\pi^0$ and one proton in the final state.
              \item $\SI{1.5}{~GeV/c} < p_\mu < \SI{20}{~GeV/c}, \, \theta_\mu < 25^\circ$.
              \item $p_\text{p} > \SI{0.45}{~GeV/c}$.
          \end{enumerate}
    \item For CC-$0\pi$~\cite{MINERvA:2018hba}:
          \begin{enumerate}
              \item $\nu_\mu + \textrm{A} \rightarrow \mu^- + \text{p} + \textrm{X}$, requiring one $\mu$ and at least one proton in the final state, with no pions.
              \item $\SI{1.5}{~GeV/c} < p_\mu < \SI{10}{~GeV/c}, \, \theta_\mu < 20^\circ$.
              \item $\SI{0.45}{~GeV/c} < p_\text{p} < \SI{1.2}{~GeV/c}, \, \theta_p < 70^\circ$.
          \end{enumerate}
\end{itemize}

For the $0\pi$ measurement, the contribution from pion production is relatively minor: arising from pion absorption, a process addressed by the FSI mechanism included in the \nuwro cascade model.
Therefore, the difference between the predictions of the $\Delta$ and Hybrid model can be anticipated to be minimal.
On the other hand,  in the $\pi^0$ measurement, there is no restriction based on $W_\textrm{rest}$.
The predicted $W$ distributions are shown in Fig.~\ref{fig:TKIW}b.
The presence of a resonance peak near $W = 1.5~\textrm{GeV}$ and the ability to extend to higher $W$ values in the Hybrid model significantly alter the distributions.
The contribution from the $\Delta$ resonance in the $\Delta$-P model ceases at $W = 1.6~\textrm{GeV}$. As explained before (Sec.~\ref{sec:hynuwro}), an apparent presence of a contribution from the second resonance region in the $\Delta$-P model results is an artifact arising from the RES-DIS transition region.  In contrast, the Hybrid model exhibits a pronounced peak in the second resonance region, which is not present in the $\Delta$ model. As a consequence, \pythia is extrapolated down to a low $W$-value in $\Delta$-P, resulting in a less realistic picture, with significantly more strength in the second resonance region compared to the H-P model.

Figure~\ref{fig:NuWro_LFG_TKI} illustrates the comparison between the predictions of the $\Delta$ and Hybrid models and the MINERvA $0\pi$ and $\pi^0$ measurements.
As mentioned before, in the $0\pi$ measurement, pion production contributes to the event sample via pion absorption during FSI (upper panels), therefore, the change of pion production model has only limited impact on the $\delta\alpha_\textrm{T}$ and $p_\textrm{N}$ distributions.
Conversely, when we compare to the $\pi^0$ measurement (lower panels in Fig.~\ref{fig:NuWro_LFG_TKI}), the model predictions become noticeably different.

\begin{figure}[!t]
    \centering
    \includegraphics[width=\figwid\textwidth]{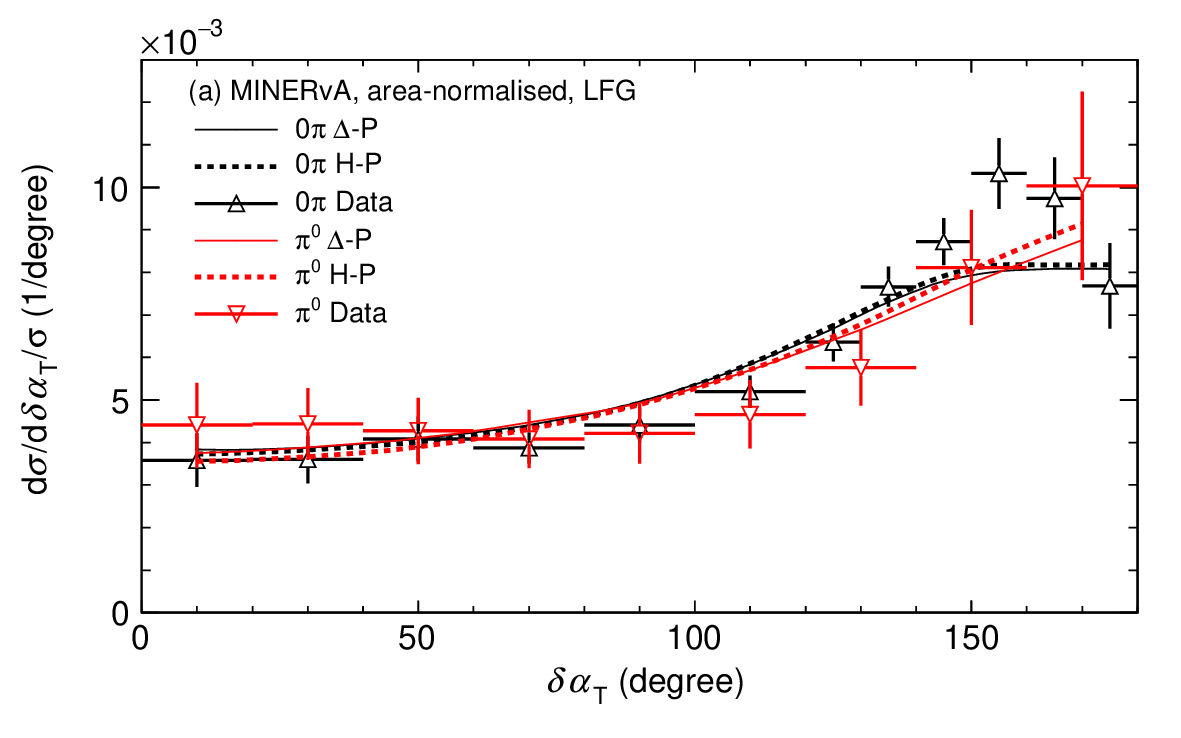}
    \includegraphics[width=\figwid\textwidth]{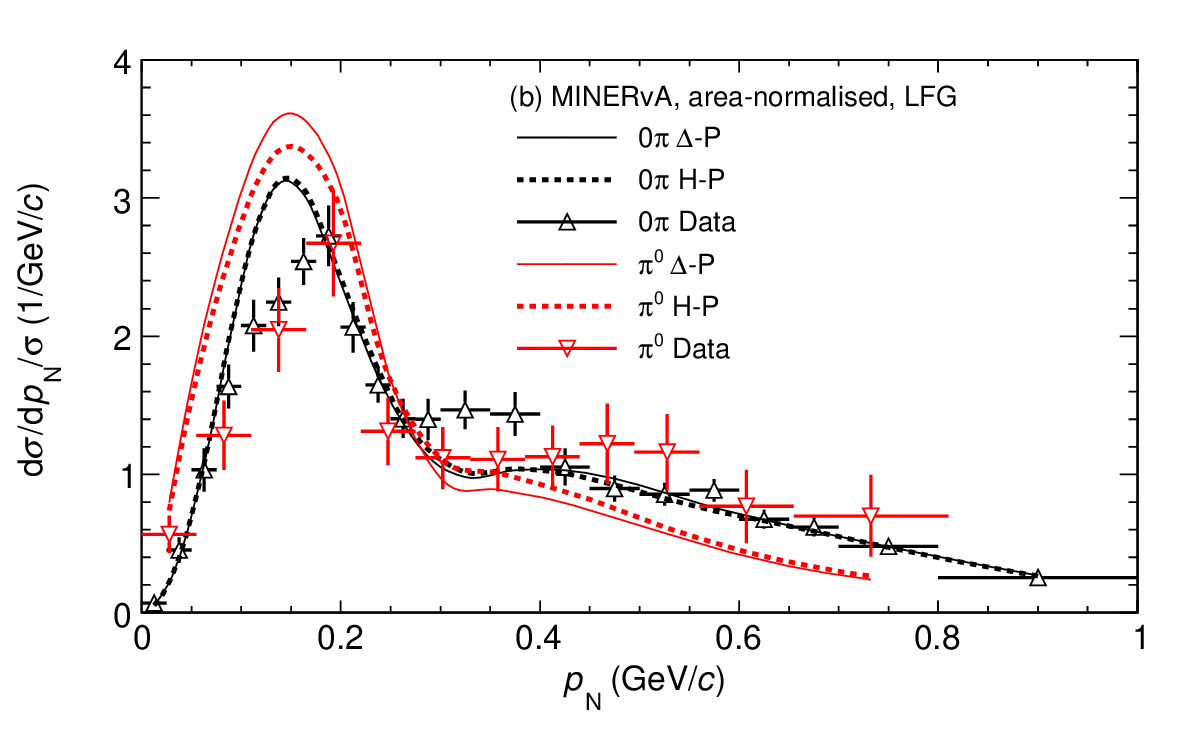}
    \caption{Shape-only comparison summarising Fig.~\ref{fig:NuWro_LFG_TKI}. Recall that LFG is only applied to non-QE channels; the initial state for QE is modelled by SF. }
    \label{fig:Hybrid_LFG_TKI_SHAPE}
\end{figure}

Firstly, a change in the normalisation is observed in $\delta\alpha_\textrm{T}$. Since the shape of $\delta\alpha_\textrm{T}$ is dictated by FSI, which is independent of the pion production modeling, we do not expect to see shape differences for the $\Delta$ and Hybrid models predictions.
Note that, as was pointed out in the original paper~\cite{MINERvA:2020anu}, it is interesting that the $\delta\alpha_\textrm{T}$ shape from the two measurements happens to be similar (Fig.~\ref{fig:Hybrid_LFG_TKI_SHAPE}a). Now, this similarity is also captured by the models.

\begin{figure}[tb!]
    \centering
    \includegraphics[width=\figwid\textwidth]{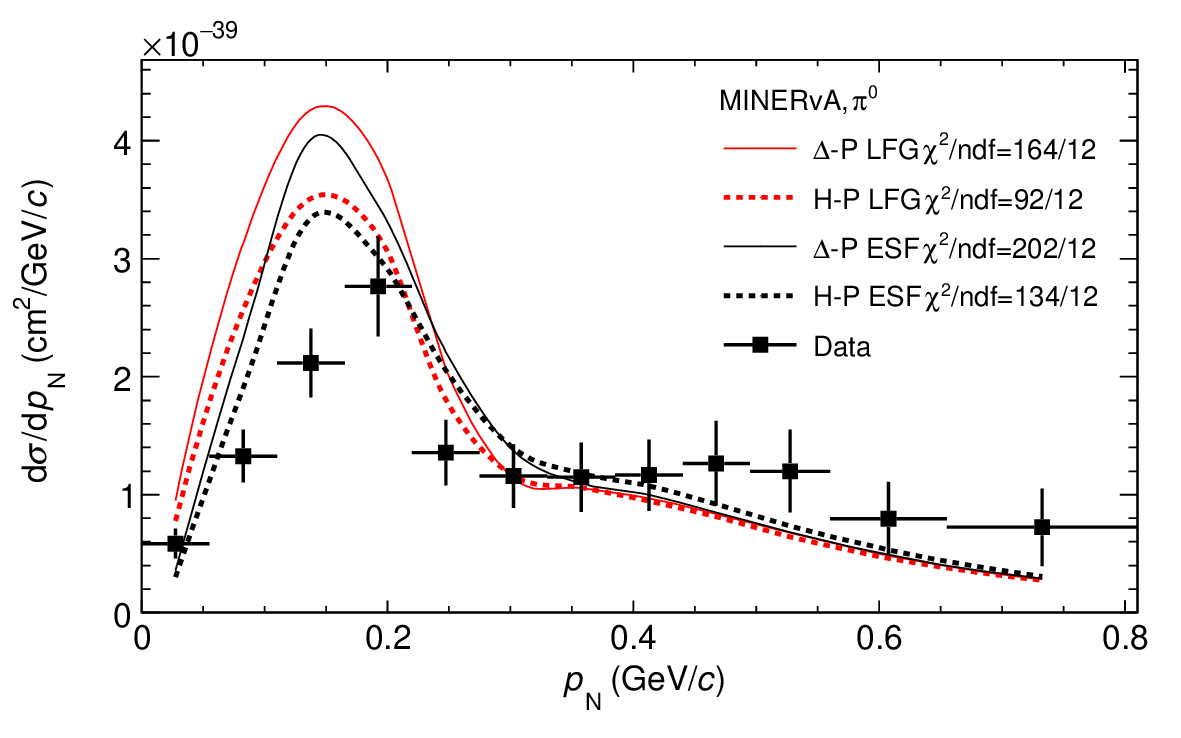}
    \caption{Additional comparison to $p_\textrm{N}$ with the initial state ESF. }
    \label{fig:NuWro_ESF_TKI1}
\end{figure}

Secondly, in the $\pi^0$ measurement depicted in Fig.~\ref{fig:NuWro_LFG_TKI}, the Hybrid model's prediction for the $p_\textrm{N}$ distribution exhibits a notable reduction in the Fermi motion peak compared to the $\Delta$ model, leading to better agreement with the data.
It is noteworthy that the $p_\textrm{N}$ shape is similar between the $0\pi$ and $\pi^0$ measurements (Fig.~\ref{fig:Hybrid_LFG_TKI_SHAPE}b).
However, despite the improvement over the $\Delta$ model, the Hybrid model still faces challenges in fully capturing the shape of the Fermi motion peak.
Further insight is gained by replacing the initial state of LFG with ESF for the $\pi^0$ measurement (see Appendix~\ref{sec:appESF} for full comparison plots with ESF).
As shown in Fig.~\ref{fig:NuWro_ESF_TKI1}, the Hybrid model also shows improvement with ESF.
However, while ESF results in a lower peak compared to LFG, it yields a higher $\chi^2$, indicating that ESF also struggles to accurately capture the shape of the Fermi motion peak.

\begin{figure}[!htb]
    \centering
    \includegraphics[width=\figwid\textwidth]{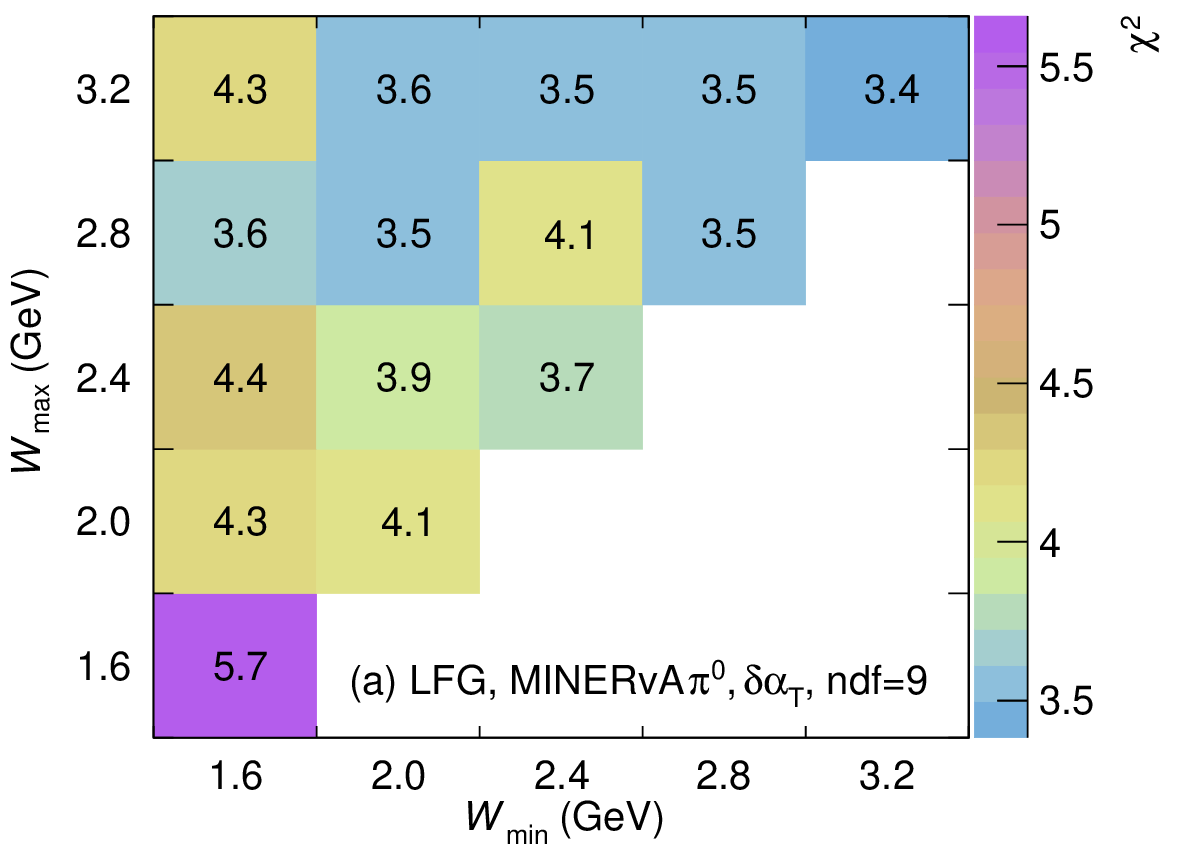}
    \includegraphics[width=\figwid\textwidth]{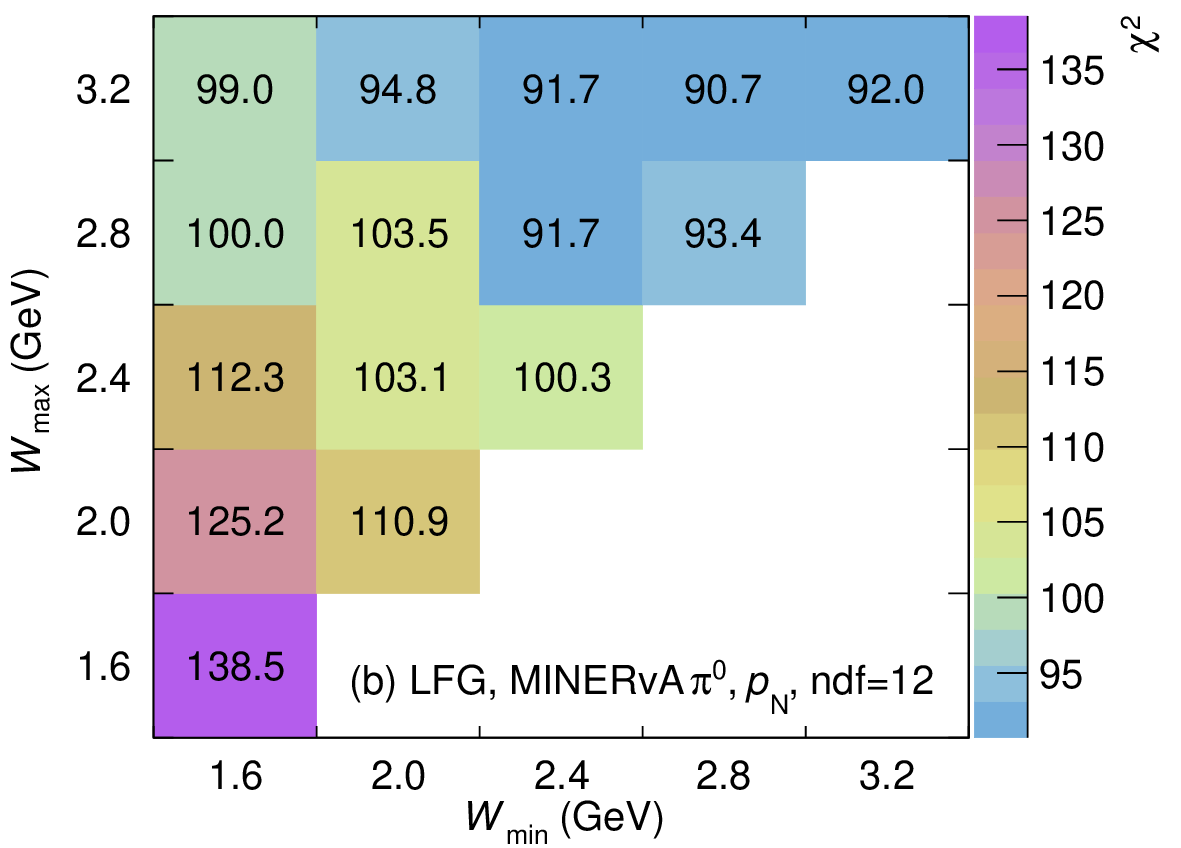}
    \caption{The $\chi^2$ value of the H-P predictions on the MINERvA $\pi^0$ measurement with varying transition window, $(W_\textrm{min},~W_\textrm{max})$.}
    \label{fig:chi2scan}
\end{figure}

\begin{figure}[!htb]
    \centering
    \includegraphics[width=\figwid\textwidth]{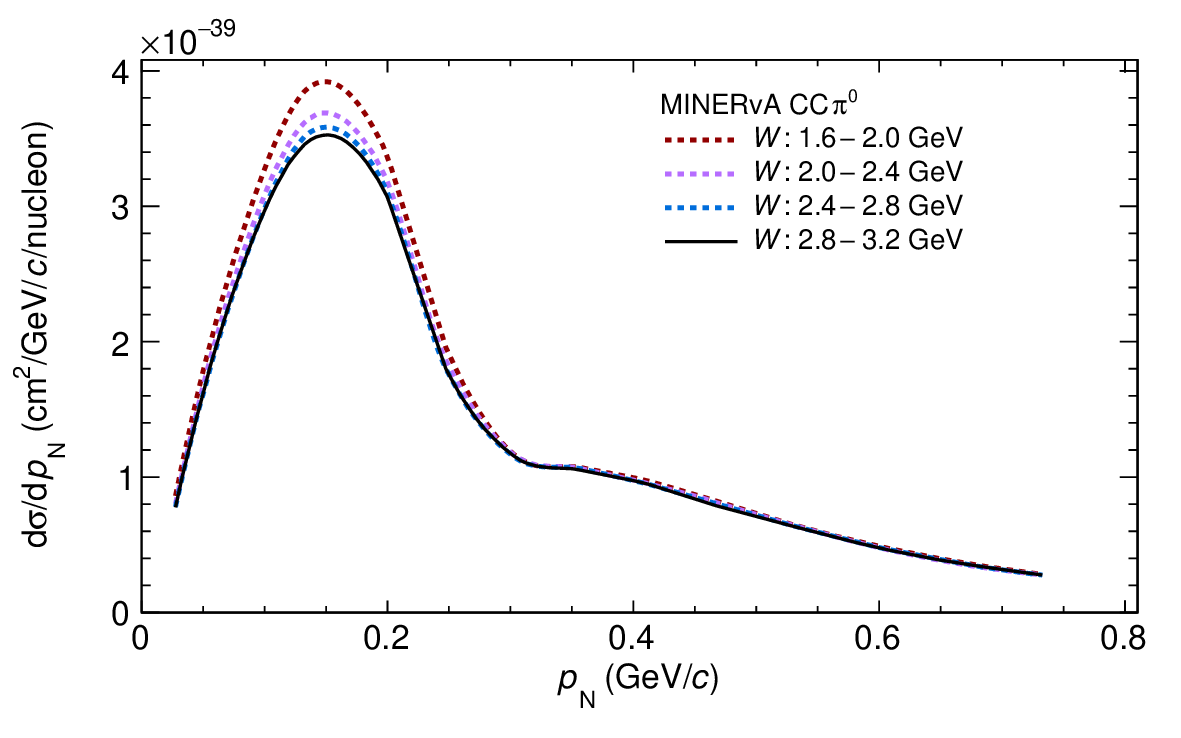}
    \caption{The H-P model prediction of $\pn$ distribution for MINERvA CC$\pi^0$ with different transition regions. The 2.8-3.2~GeV window is the nominal choice of our model.}
    \label{fig:pn_vary_region}
\end{figure}

Part of the improvement observed in the Hybrid model indeed arises from its refined depiction of interactions in higher $W$ regions. This can be demonstrated by varying the starting point ($W_\textrm{min}$) and the stopping point ($W_\textrm{max}$) of the H-P transition region (Fig.~\ref{fig:chi2scan}).
Adjusting the transition window, $(W_\textrm{min},~W_\textrm{max})$, in the H-P model to the higher end gives more phase space to the Hybrid model and less to \pythia.
Clear trends are observed in Fig.~\ref{fig:chi2scan}: as both $W_\textrm{min}$ and $W_\textrm{max}$ increase, the $\chi^2$ decreases, indicating that the Hybrid model provides an improved description of the data compared to \pythia, in the kinematic region of interest for the MINERvA $\pi^0$ TKI analysis.
When the transition begins at a sufficiently high $W$ ($> \SI{2.4}{GeV}$), the model consistently yields an optimized chi-squared value.
This is consistent with the $\dv{\sigma}{W}$ results in Fig.~\ref{fig:TKIW}, which indicate that the transition primarily occurs in the high-$W$ region, thereby minimizing \pythia's influence on the final prediction.
An intuitive illustration of this effect is shown in Fig.~\ref{fig:pn_vary_region}, where $\dv{\sigma}{\pn}$ is calculated for a fixed transition width of $\SI{0.4}{GeV}$ at various $W$ values.  As $W$ increases, the cross section decreases. In this particular case, it tends to
converge to the prediction by the pure Hybrid mode.
This reduced Fermi peak contributes to a more accurate description of the data.

\section{Conclusions}\label{conclusion}

Neutrino pion production is an important interaction mechanism at GeV neutrino facilities.
In this work, we describe the implementation of the Ghent single-pion production (SPP) model~\cite{Gonzalez-Jimenez17} in the neutrino event generator \nuwro.
Unlike the  $\Delta$-\pythia ($\Delta$-P) implementation, until now used in \nuwro, the resulting Hybrid-\pythia (H-P) model essentially removes
contributions from \pythia to the SPP  channel. The description of SPP is improved through extra resonance contributions in the second resonance region, interference with the non-resonant background of Ref.~\cite{Hernandez07}, and a Regge description at high $W$.

Predictions by the H-P model are compared to MINERvA and MicrooBooNE (and T2K, see Appendix~\ref{sec:appT2K}) data.
As expected, the refined modeling in regions of higher $W$ tend to improve the agreement with the data. Notably, in comparison to the TKI measurements, the overprediction in the $p_\textrm{N}$ Fermi-motion peak region, as previously seen from the $\Delta$-P model, has now been significantly reduced, indicating an improvement at the pion production level---the remaining overestimate, however, could be ambiguously attributed to either the production cross section or FSIs~\cite{GENIE:2024ufm}, as events can migrate from the $\pn$ peak to the tail by modifying the latter.
The shapes of the $\delta\alpha_\textrm{T}$ and $p_\textrm{N}$ distributions indicate some universality between pionless and pion production; however, capturing the shape of the $p_\textrm{N}$ Fermi-motion peak  turns out to be a challenge for current nuclear models implemented in MC generators.
This observation should motivate further development of the initial state modelling for pion production, beyond the local Fermi gas and the effective spectral function approaches.

Modification of the pion production mechanism due to in-medium effects may also lead to important changes in the predicted cross sections~\cite{bogart2024inmedium}. Medium modification of the $\Delta$ resonance~\cite{OSET1987631} are not explicitly included here and can lead to a non-negligible reduction of the cross section for MINERvA kinematics~\cite{Gonzalez-Jimenez18}.
This is the case of the modification of the $\Delta$ decay width, which in the model of~\cite{OSET1987631} leads to the opening of pionless $\Delta$ decay channels. This mechanism is partially included in the \nuwro cascade as pion absorption;
however, further and dedicated studies on this subject are needed.

Apart from nucleonic effects, there are nuclear effects that should be addressed when comparing with neutrino-nucleus cross section data. In this paper, we have discussed FSI included in the intranuclear cascade and, briefly, the medium-modification of the resonance properties. However, a more comprehensive understanding should incorporate other quantum mechanical effects not addressed here, like the distortion of the hadrons (i.e., elastic FSI for the outgoing nucleon and pion), which cannot be addressed by classical or semiclassical approaches like intranuclear cascade or models based on a factorized cross section; alternatively, these effects could be incorporated into the model describing the elementary vertex~\cite{Nikolakopoulos22}.

Currently, work is in progress to improve the Hybrid model by unitarizing the amplitude and incorporating additional contributions to the SPP mechanism, such as $\rho$- and $\omega$-meson exchanged and higher mass resonances. The present work paves the road so that these and other potential improvements could be easily incorporated in \nuwro.

In the data-model comparison, we also appreciate future measurements  that expand the phase space to higher~$W$ which is crucial to differentiate pion production models in the transitional region.

\begin{acknowledgments}

    We thank Clarence Wret for pointing out an update of the MINERvA data of Ref.~\cite{MINERvA:2014ogb}.
    R.G.-J. is supported by Project No.~PID2021-127098NA-I00 funded by MCIN\slash AEI\slash 10.13039\slash 501100011033\slash FEDER, UE.
    N.J. and K.N. are supported by the Fund for Scientific Research Flanders (FWO) and by Ghent University Special Research Fund.
    X.L. is supported by the STFC (UK) Grant No.~ST\slash S003533\slash 2.
    A.N. is supported by Fermi Research Alliance, LLC under Contract No. DE-AC02-07CH11359 with the U.S. Department of Energy, Office of Science, Office of High Energy Physics.
    J.T.S. is supported by Polish Ministery of Science grant UMOWA 2023\slash WK\slash 04.
    Q.Y. and Y.Z. are supported by National Natural Science Foundation of China (NSFC) under contract 12221005.

\end{acknowledgments}

\bibliographystyle{JHEP}
\bibliography{bibliography}

\newpage

\appendix

\section{Comparison with T2K $\pi^+$ TKI data}\label{sec:appT2K}
With a neutrino beam with an energy of approximately 0.6~GeV scattering off a hydrocarbon target, T2K has measured TKI in the CC-$\pi^+$ production~\cite{T2K:2021naz}: $\nu_\mu + \textrm{A} \rightarrow \mu^- + \text{p} + \pi^+ + \textrm{X}$, where X is the hadronic system that can contain nucleons but no mesons.  The phase space cuts are defined in Table~\ref{tab:T2K_TKI}.

\begin{table}[!htb]
    \centering
    \begin{tabular}{lcr}
        \hline
        Particle & Momentum        $p$                 & Angle $\theta$ \\\hline
        $\mu^-$  & $\qtyrange{0.25}{7}{~GeV/\it{c}}$   & $<70 \degree$  \\
        $\pi^+$  & $\qtyrange{0.15}{1.2}{~GeV/\it{c}}$ & $<70 \degree$  \\
        p        & $\qtyrange{0.45}{1.2}{~GeV/\it{c}}$ & $<70 \degree$  \\
        \hline
    \end{tabular}
    \caption{The phase space cuts definition for T2K CC-$\pi^+$ TKI measurement~\cite{T2K:2021naz}.}\label{tab:T2K_TKI}
\end{table}

\begin{figure}[!htb]
    \centering
    \includegraphics[width=\figwid\textwidth]{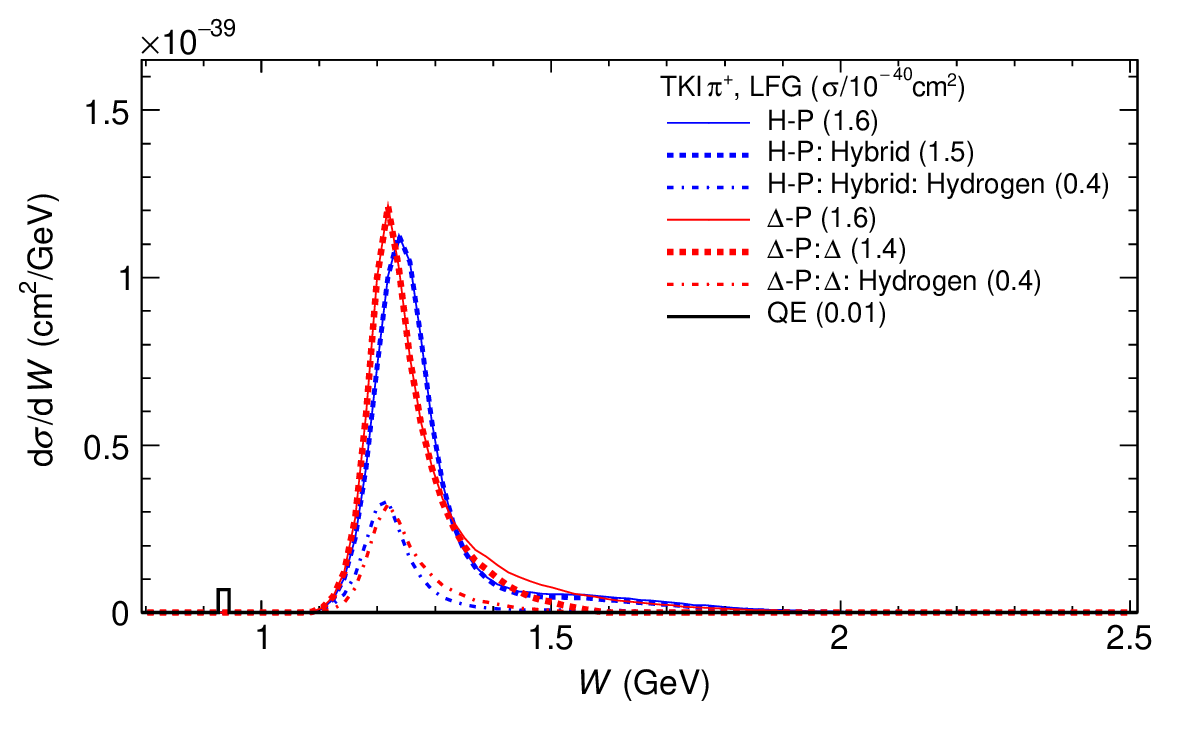}
    \caption{Similar to Fig.~\ref{fig:mpiW} but for the T2K $\pi^+$ TKI  measurement~\cite{T2K:2021naz}. }
    \label{fig:T2K_TKI_W}
\end{figure}

\begin{figure*}[!htb]
    \centering
    \includegraphics[width=\twofigwid\textwidth]{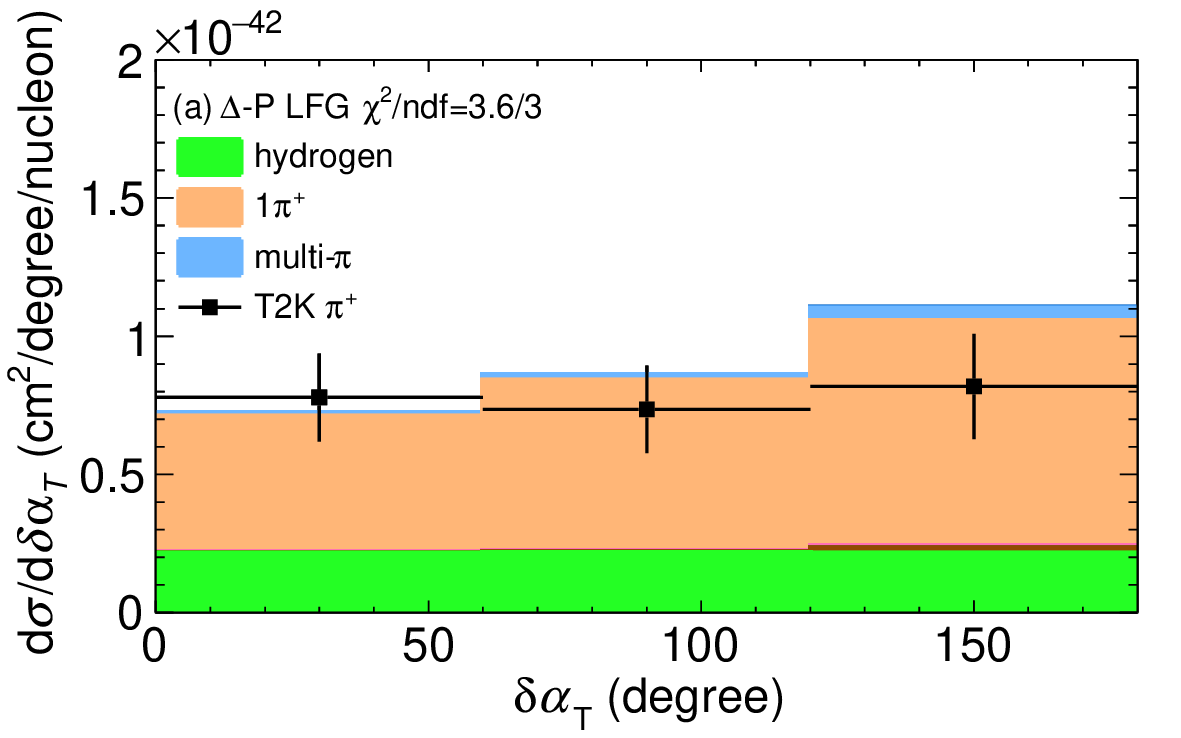}
    \includegraphics[width=\twofigwid\textwidth]{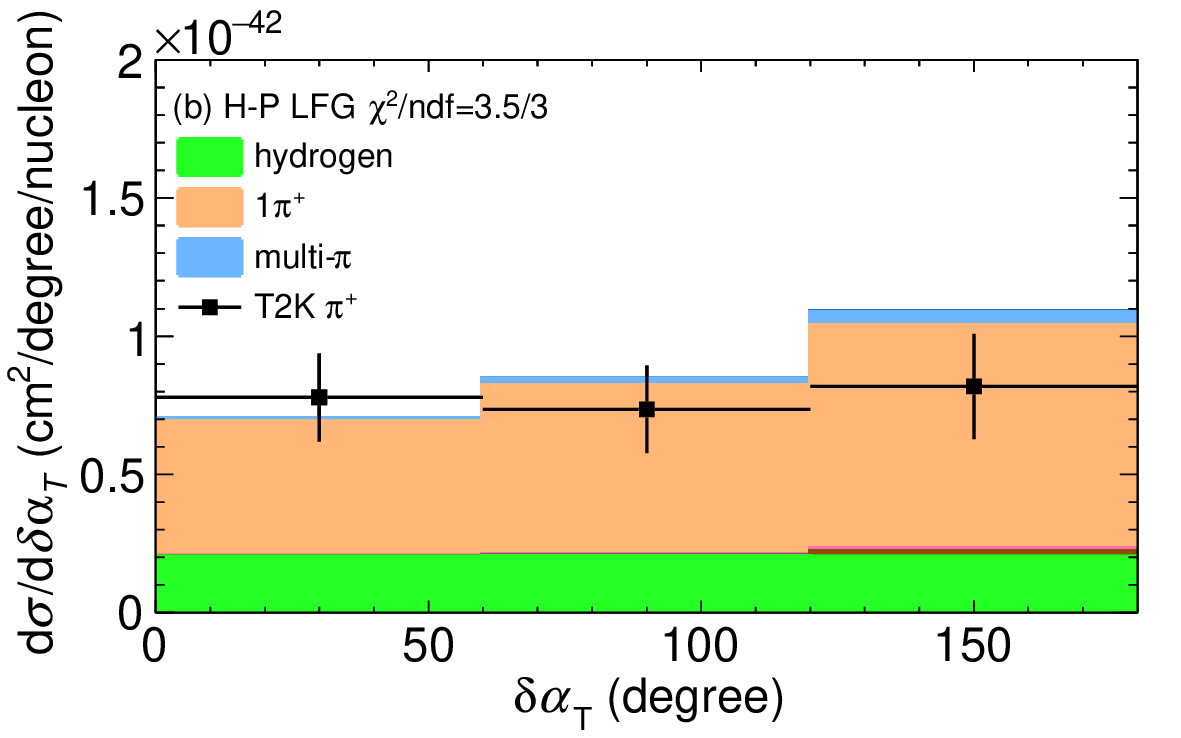}
    \includegraphics[width=\twofigwid\textwidth]{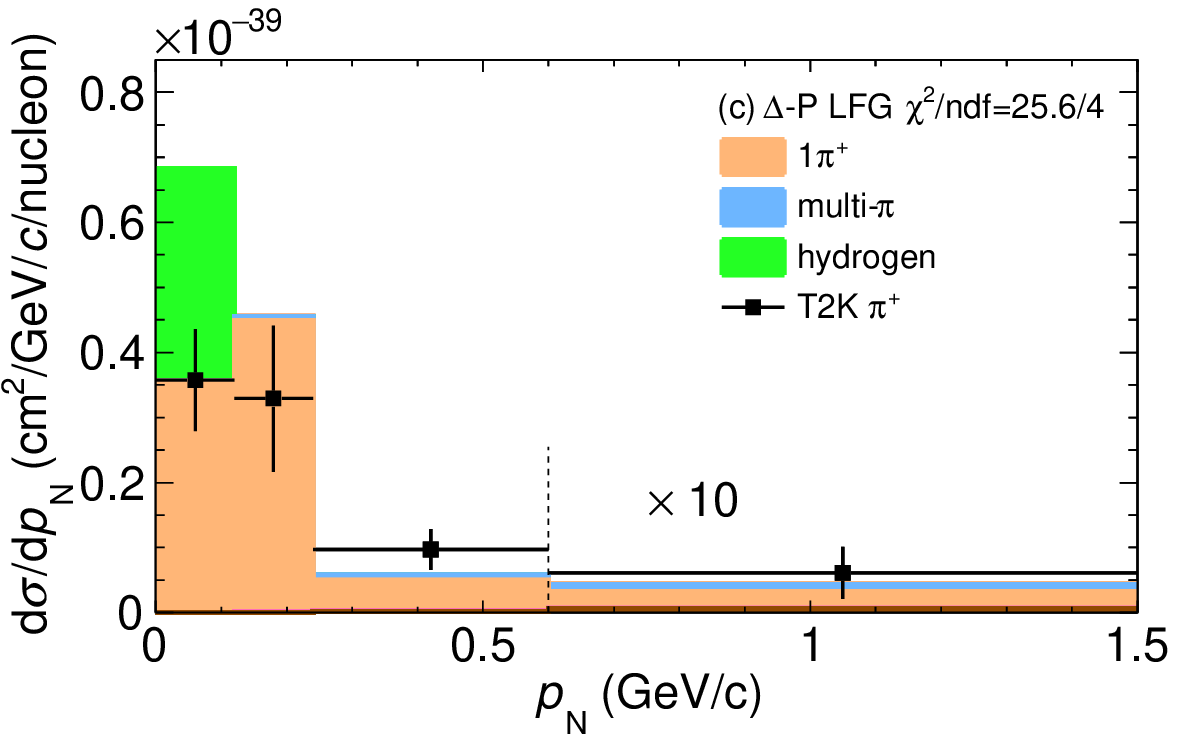}
    \includegraphics[width=\twofigwid\textwidth]{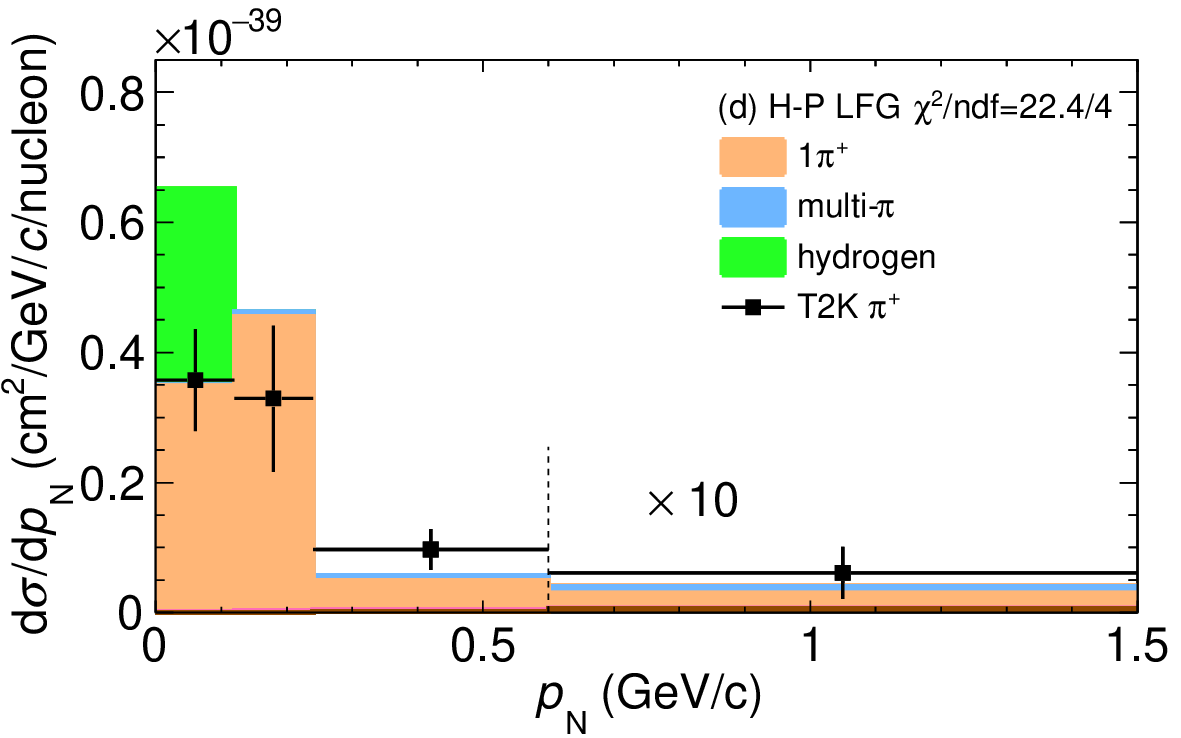}
    \caption{Similar to Fig.~\ref{fig:NuWro_LFG_TKI} lower panels but for the T2K  $\pi^+$ TKI measurement~\cite{T2K:2021naz}.}
    \label{fig:T2K_TKI}
\end{figure*}

As shown in Fig.~\ref{fig:T2K_TKI_W}, the $W$-distribution is dominated by the $\Delta^{++}$ resonance, as  expected given the low energy of the neutrinos. Figure~\ref{fig:T2K_TKI} depicts the $\Delta$-P and H-P predictions, showing limited improvement by the latter due to the low-$W$ dominance of the sample. As noted by Ref.~\cite{T2K:2021naz}, the inherent challenge in describing the data appears to stem from the initial state of LFG.

\section{Comparison with MINERvA TKI data using ESF}\label{sec:appESF}

\begin{figure*}[!htb]
    \centering
    \includegraphics[width=\twofigwid\textwidth]{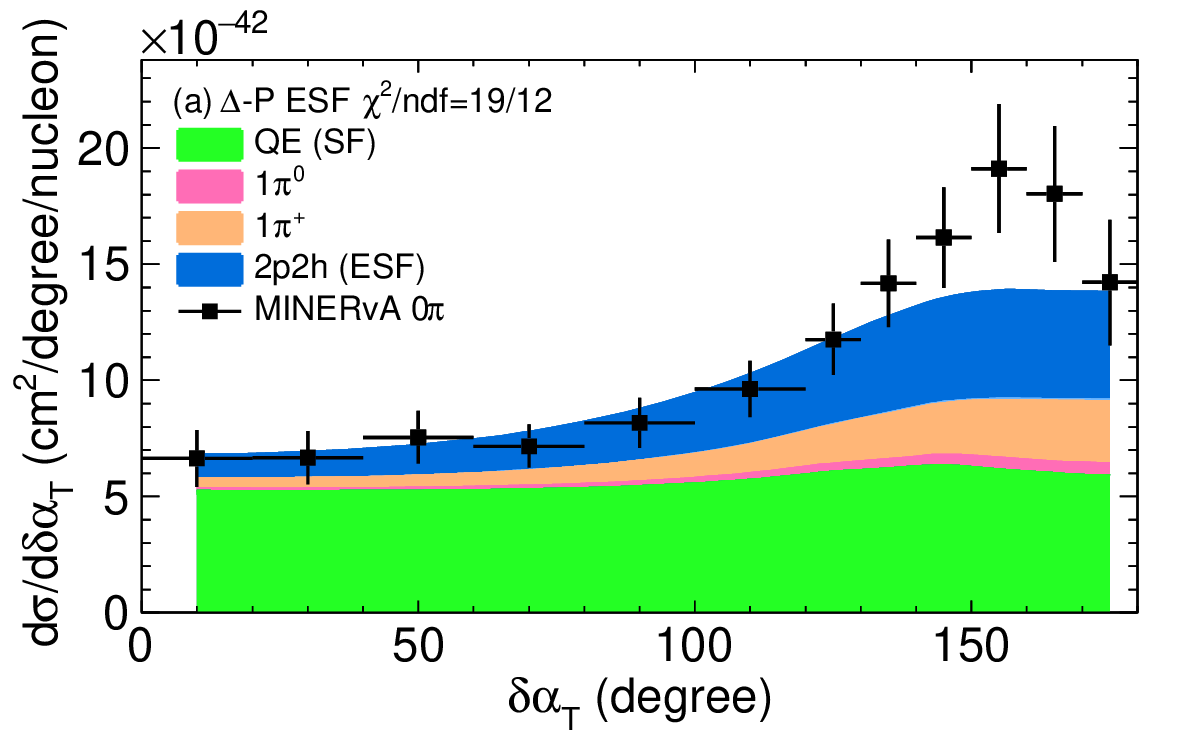}
    \includegraphics[width=\twofigwid\textwidth]{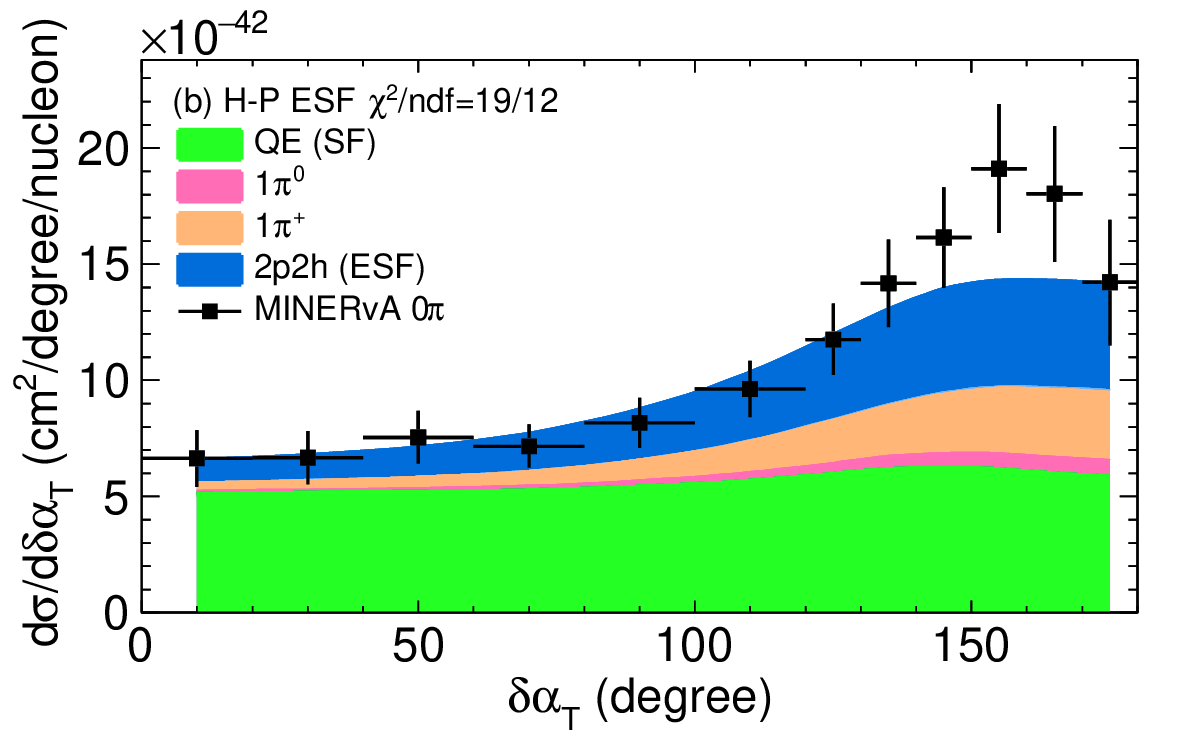}

    \includegraphics[width=\twofigwid\textwidth]{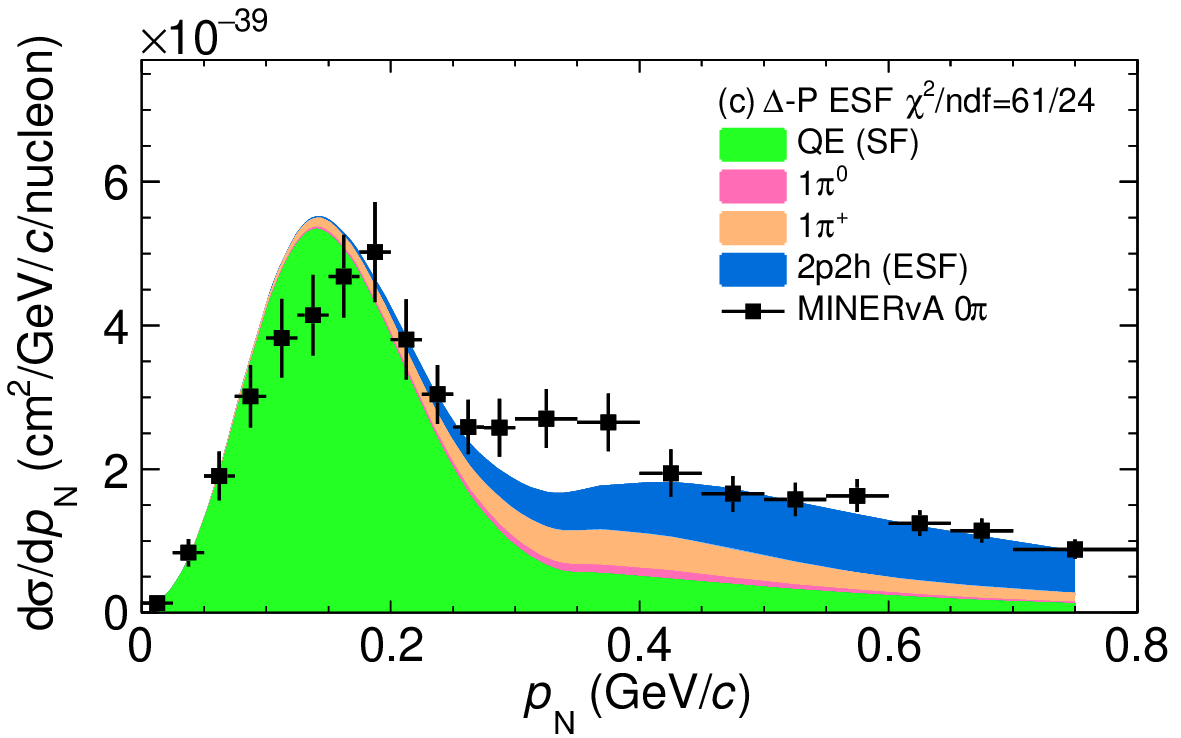}
    \includegraphics[width=\twofigwid\textwidth]{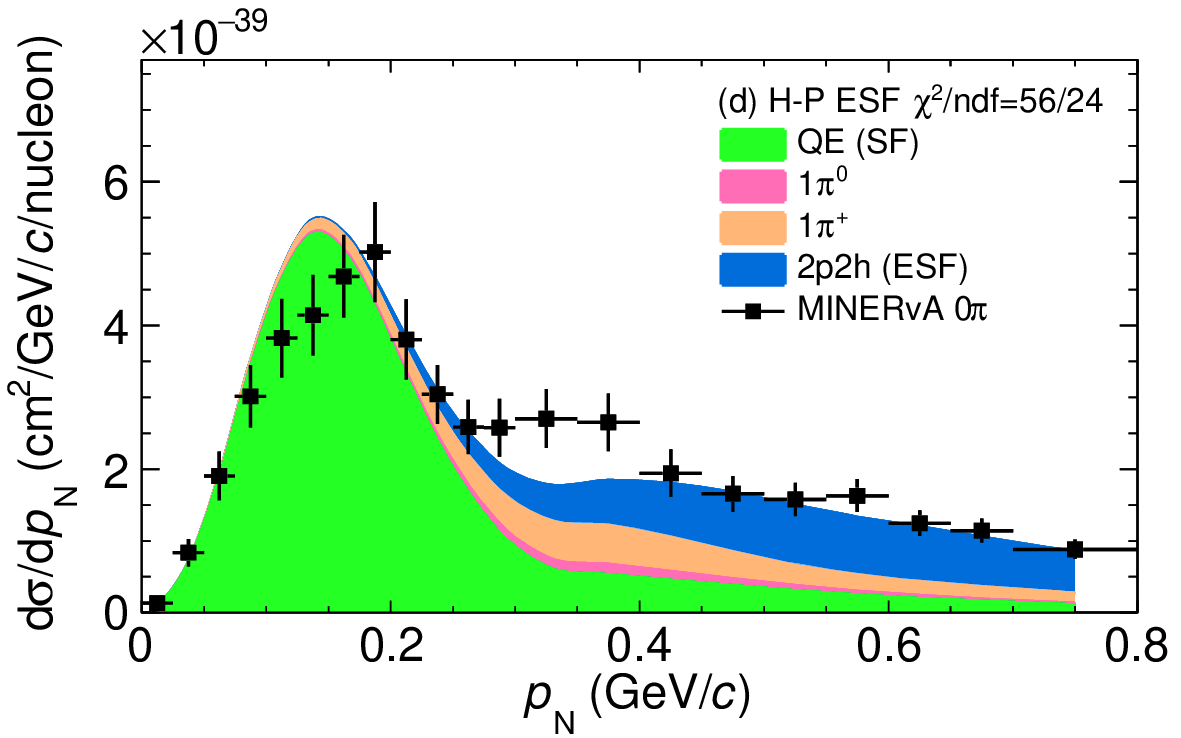}

    \includegraphics[width=\twofigwid\textwidth]{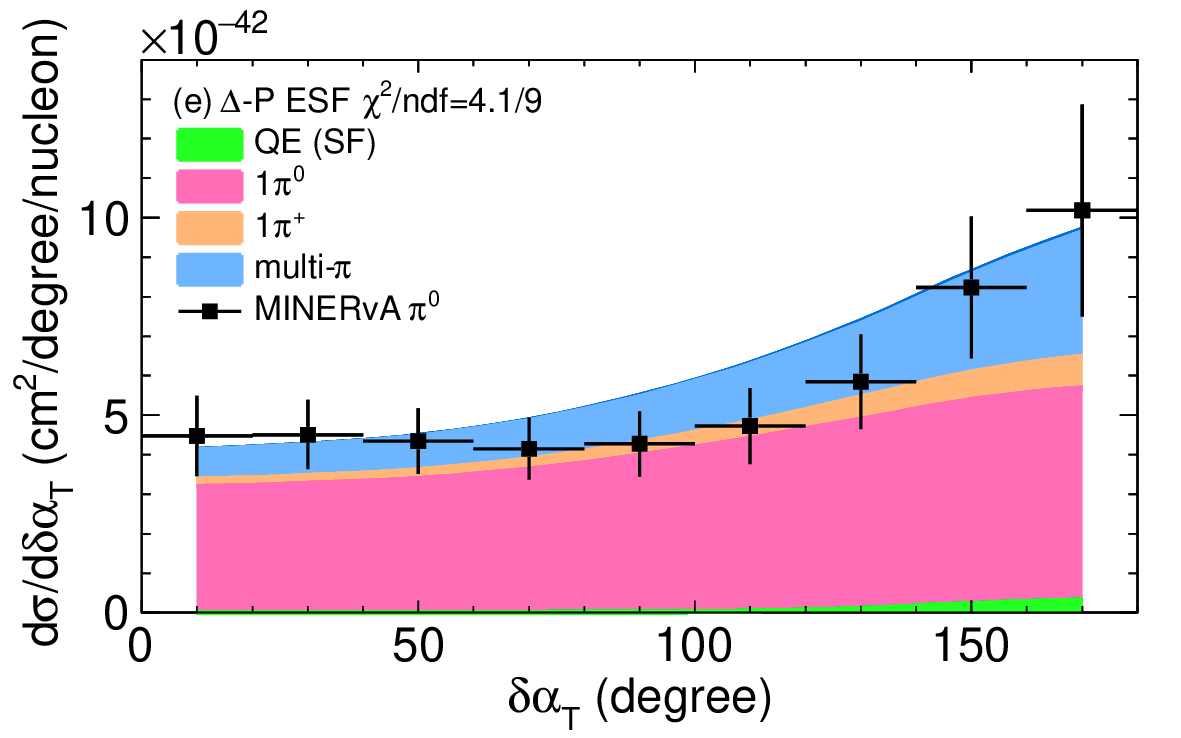}
    \includegraphics[width=\twofigwid\textwidth]{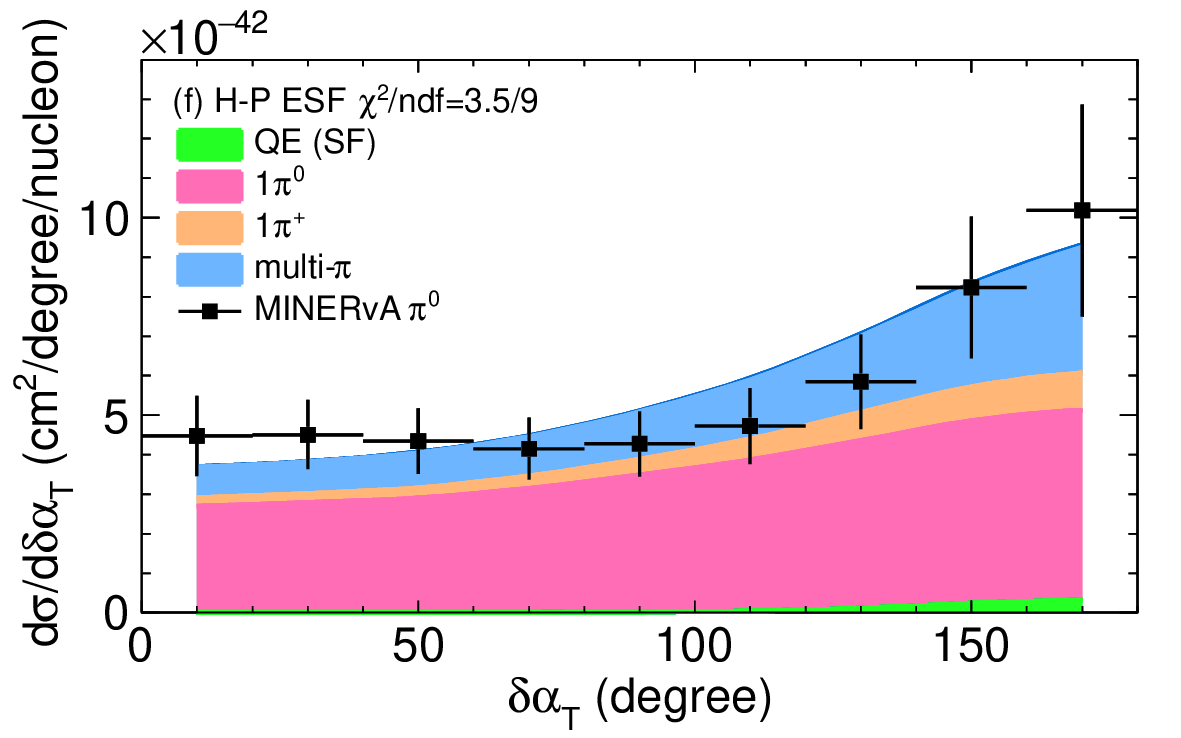}

    \includegraphics[width=\twofigwid\textwidth]{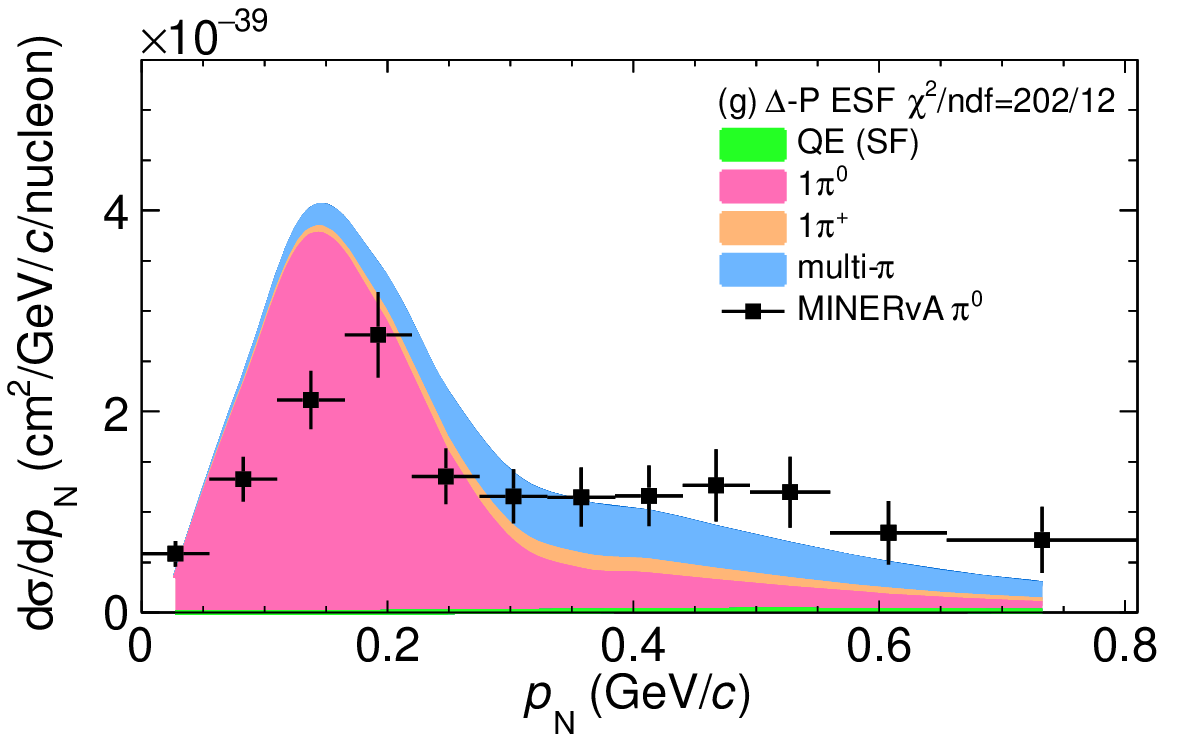}
    \includegraphics[width=\twofigwid\textwidth]{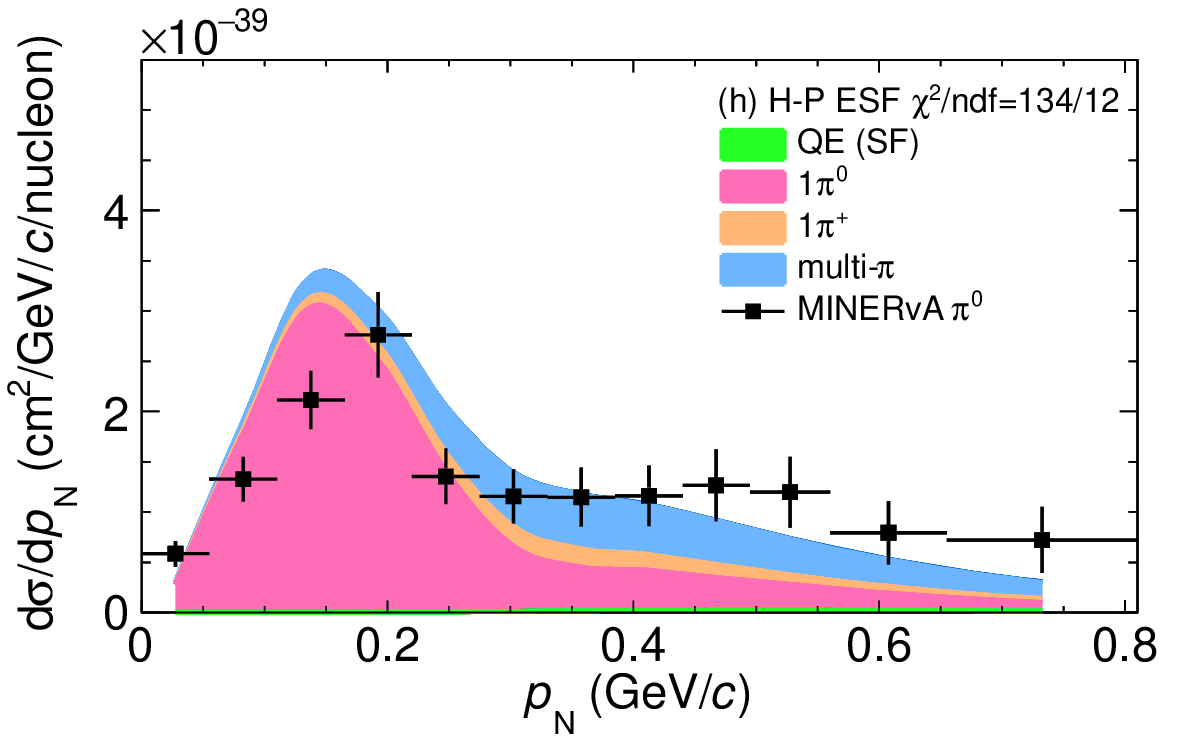}

    \caption{Similar to Fig.~\ref{fig:NuWro_LFG_TKI} but with the non-QE initial state modelled by ESF.}
    \label{fig:NuWro_ESF_TKI}
\end{figure*}

For completeness, the comparison with the MINERvA data using the initial state ESF is fully shown in Fig.~\ref{fig:NuWro_ESF_TKI}.

\section{Single pion production $\dv{\sigma}{W}$ with 3~GeV neutrino}\label{sec:appSPP}
\begin{figure}[!htb]
    \centering
    \includegraphics[width=\figwid\textwidth]{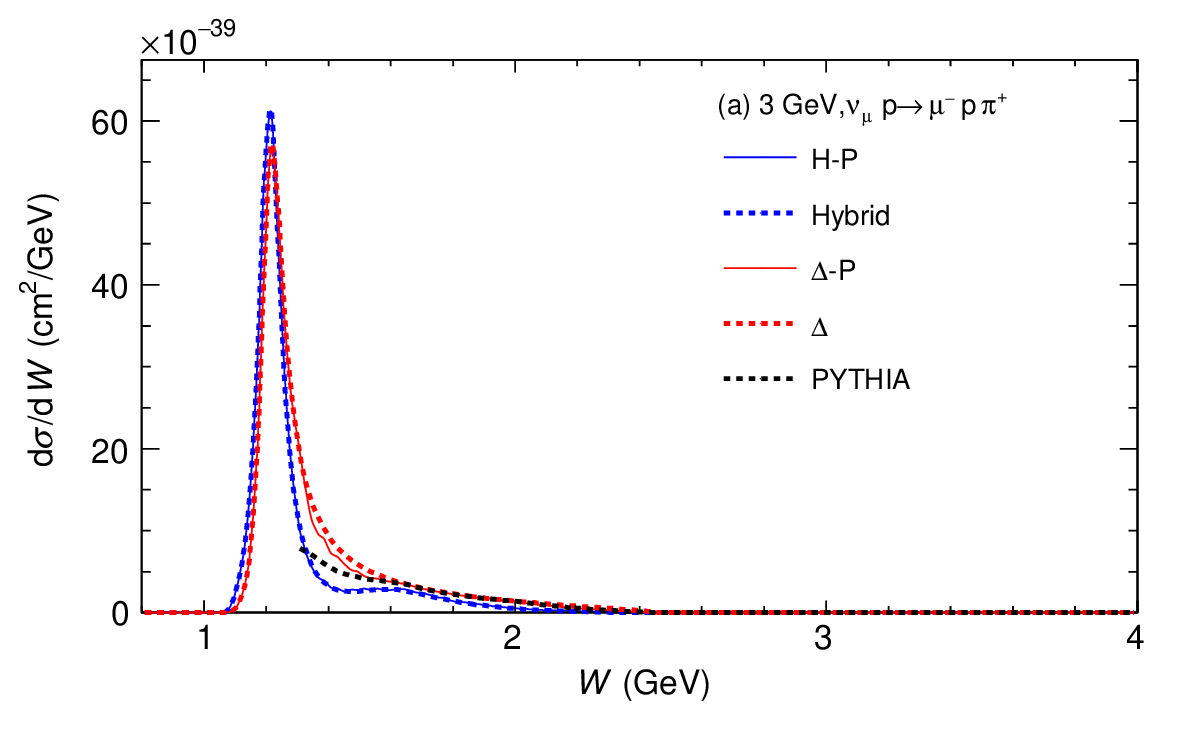}
    \includegraphics[width=\figwid\textwidth]{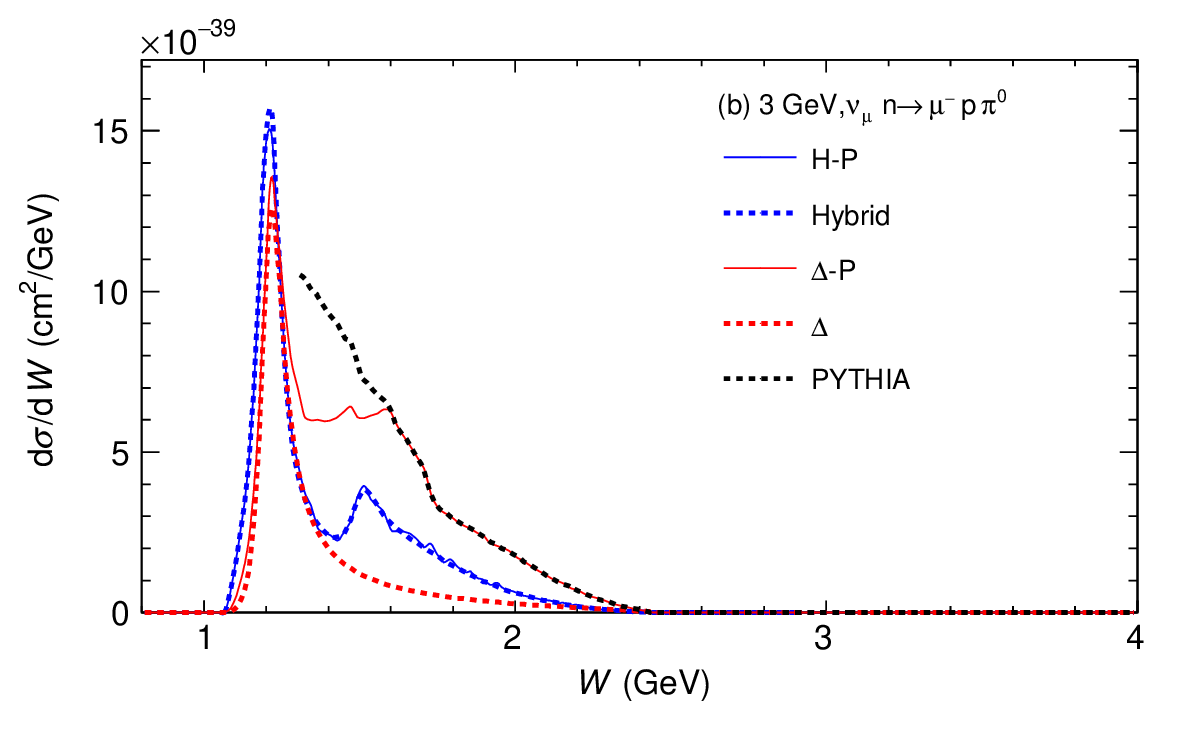}
    \includegraphics[width=\figwid\textwidth]{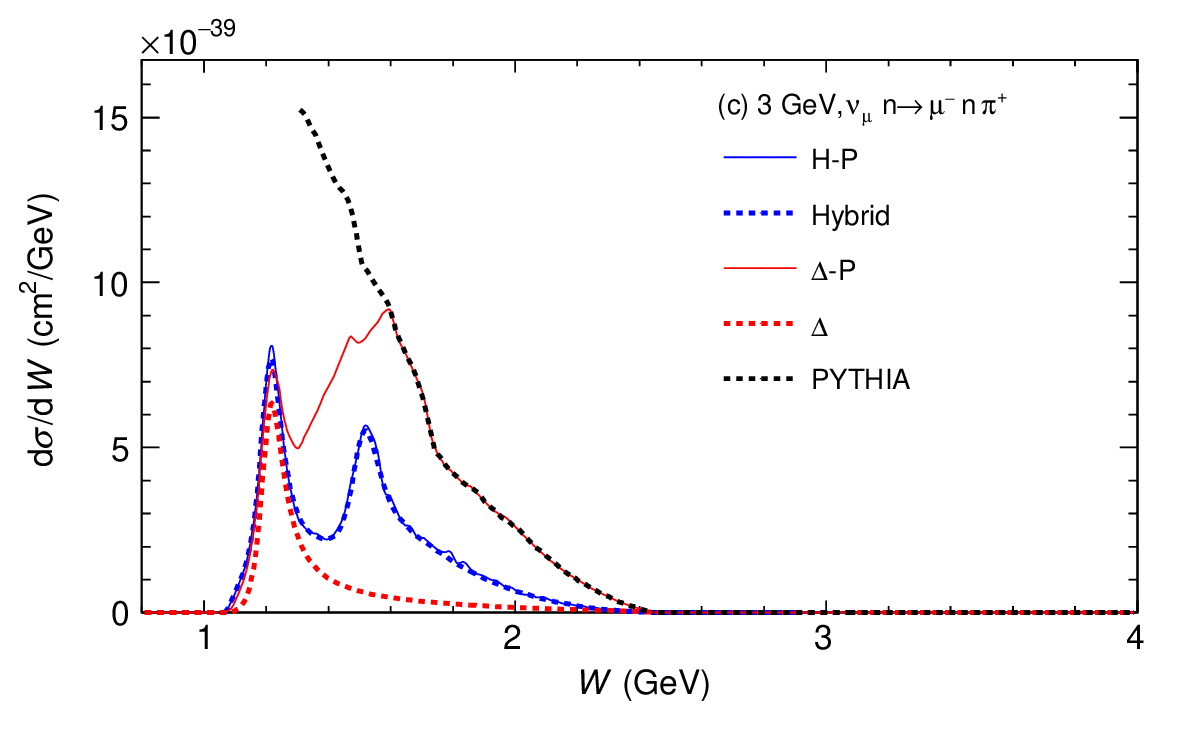}
    \caption{$\Delta$ and Hybrid model predictions for SPP by an 3~GeV neutrino off the nucleon: (a) $\nu_\mu\textrm{p}\rightarrow\mu^-\textrm{p}\pi^+$, (b) $\nu_\mu\textrm{n}\rightarrow\mu^-\textrm{p}\pi^0$, and (c) $\nu_\mu\textrm{n}\rightarrow\mu^-\textrm{n}\pi^+$. The trend and difference between models are similar to the figure with 8~GeV neutrino in Fig.~\ref{fig:trans}. But the lower energy suppressed the contribution in the transition region.
    }
    \label{fig:3GeVtrans}
\end{figure}

To illustrate model behavior, a neutrino energy of $\SI{8}{GeV}$ is used in Fig. \ref{fig:trans}. For comparison, Figure~\ref{fig:3GeVtrans} presents the same analysis for a lower energy of $\SI{3}{GeV}$. As can be seen, at $\SI{3}{GeV}$, the contribution from the transition region is diminished.

\end{document}